\documentclass[10pt,letterpaper]{article}
\usepackage[top=0.85in,left=1.5in,footskip=0.75in]{geometry}

\usepackage{amsmath,amssymb}

\usepackage{changepage}

\usepackage[utf8x]{inputenc}

\usepackage{textcomp,marvosym}

\usepackage{cite}

\usepackage{nameref,hyperref}

\usepackage[right]{lineno}

\usepackage{microtype}
\DisableLigatures[f]{encoding = *, family = * }

\usepackage[table]{xcolor}

\usepackage{array}

\newcolumntype{+}{!{\vrule width 2pt}}

\newlength\savedwidth

\raggedright
\usepackage[aboveskip=1pt,labelfont=bf,labelsep=period,justification=raggedright,singlelinecheck=off]{caption}

\makeatletter
\renewcommand{\@biblabel}[1]{\quad#1.}
\makeatother

\usepackage{lastpage,fancyhdr,graphicx}
\usepackage{epstopdf}
\fancyheadoffset[L]{2.25in}
\fancyfootoffset[L]{2.25in}
\newcommand{\real}{\ensuremath{\mathbb{R}}}

\newcommand{\ie}{{\it i.e.}~}

\begin{document}
\vspace*{0.2in}

\begin{flushleft}
{\Large
\textbf\newline{Agent-Level Pandemic Simulation (ALPS) for Analyzing Effects of Lockdown Measures} 
}
\newline
\\
Anuj Srivastava \\ \textsuperscript
\bigskip
Department of Statistics, Florida State University, Tallahassee, FL, USA

%
%





* anuj@stat.fsu.edu

\end{flushleft}
\section*{Abstract}
This paper develops an agent-level simulation model, termed ALPS, 
for simulating the spread of an infectious disease in a 
confined community. The mechanism of transmission is agent-to-agent contact, using 
parameters reported for Corona COVID-19 pandemic. The main goal of the ALPS simulation is 
analyze effects of preventive measures -- imposition and lifting of lockdown norms -- 
on the rates of infections, fatalities and recoveries. The model assumptions and 
choices represent a balance between 
competing demands of being realistic and being efficient for real-time inferences.
The model provides quantification of gains in reducing casualties by imposition  
and maintenance of restrictive measures in place.
\\
\noindent
{\bf Keywords}: COVID-19, Corona virus, lockdown measures, social distancing, agent-level models, SIR model




\section{Introduction}
There is a great interest in statistical modeling and analysis of medical, 
economical and epidemiological data resulting from current
the Covid-19 pandemic. From an epidemiological perspective, 
as large amount of infection, containment, and recovery data from the
this pandemic becomes available over time, the 
community is currently relying essentially on simulation models to help assess situations and to evaluate options 
\cite{nature-news-adam}.
Naturally, simulation systems that follow precise mathematical and statistical models are playing an important 
role in understanding this dynamic and complex situation \cite{chao-plosCB:2010}. 
In this paper we develop a mathematical 
simulation model, termed ALPS, to replicate the spread of an infectious disease, such as COVID-19, in a 
confined community and 
to study the influence of some governmental interventions on final outcomes. 

Since ALPS is purely a simulation model, the underlying assumptions and 
choices of statistical distributions for random quantities become 
critical in its success.
On one hand it is important to capture the intricacies
the observed phenomena as closely as possible, using sophisticated modeling tools. 
On the other hand, it is important to keep the model efficient and tractable by using simplifying assumptions.
One can, of course, 
relax these assumptions and obtain more and more realistic models as desired, but at the cost of 
increasing computational complexity.

There have been a large models proposed in the literatures relating to the 
the spread of epidemics through human contacts or otherwise. They can be broadly 
categorized in two main classes (a more detailed taxonomy of simulation models can 
be found in \cite{hunter2017}):

\begin{enumerate}
\item {\bf Population-Level Coarse Modeling}: 
A large number of 
epidemiological models have focused on the population-level variables - counts of infected (I), 
susceptible (S), removed or recovered (R), etc. The most popular model of this type is the 
Susceptible-Infected-Removed (SIR) model \cite{SIR-model} proposed by Kermack and McKendrick 
in 1927. This model uses ordinary differential equations to model constrained growth of  
the counts in these three categories:
$$
{dS\over dt} = -\beta I(t) S(t),\ \ 
{dI \over dt} = \beta I(t) S(t) - \gamma I(t),\ \ {dR \over dt} = \gamma I(t)\ .
$$
The two parameters $\beta$ and $\gamma$ control the dynamics of infections, 
and the condition ${dS\over dt} +  {dI \over dt} + {dR \over dt} = 0$ ensures 
constancy of the community size.
A number of other papers have studied variants 
of these models and have adapted them for different epidemics, such as ebola and SARS
\cite{timpka-who-2008}. 
While there are spatial versions of SIR models, they are usually limited in their modeling of
spatial dynamics. They typically use a uniform static grid to represent the
spread of infections, from a location to its neighbors, over time. 
In general these models do not explicitly model people dynamics
as residents move around in a community. 
Several recent simulation models, focusing directly on Covid-19 illness, also rely on
such coarser community level models \cite{ferguson-Lancet-2020}. 

\item {\bf Agent-Level Modeling}: 
While population-level dynamical evolutions of population variables are simple and very effective for 
overall assessment, they do not take into account any social dynamics, human behavior, 
government-mandated restrictions, 
and complexities of human interactions explicitly. The models that study these human-level factors and variables , 
while tracking disease at an individual level, are called {\it agent-level models}
\cite{agent-models-gilbert}. Here one models 
the mobility, health status, and interactions of individual subjects (agents) in order to construct an 
overall population-level picture in a bottom-up way. The advantages of agent-based models are that they are more 
detailed and one can vary the parameters of restriction measures, 
such as social distancing, at a granular level to infer overall 
outcomes. 
Agent-based models have been discussed in several papers, including \cite{epstein-axtell,perez-etal-2009,hunter-etal-PLOS:2018,hunter2017} and so on.
The importance of simulations based analysis of epidemic spread is emphasized in 
\cite{nguyen-BMC} but with a focus on infection models within host. Some aspect of spreading 
of diseases using network contact is also discussed.  
Hunter et al. \cite{hunter-etal-PLOS:2018} construct a detailed 
agent-based model for spread of infectious diseases, taking into account population demographics and
other social conditions, 
but they do not consider countermeasures such as lockdowns in their simulations. 
A broad organization of different agent-based simulation methods have been presented 
in \cite{hunter2017}. There are numerous other papers on the topic of agent-based simulations 
for simulating spread of infections that are not referenced. 

\end{enumerate}

The main distinction of the current paper from the past 
literature is its focus on {\bf agent-level transmission of infections}, and the 
{\bf influence of social dynamics and lockdown-type restrictions} on these transmissions. 
In this paper we assume a closed community with the infection started by a 
single agent at initial time. The infections are
transmitted through physical exposure of susceptible agents to the infected agents. The infected
agents go through a period of sickness with two eventual outcomes -- full recovery for most and 
death for a small fraction. Once recovered, the agents can no longer be infected. 

The social dynamical model used here is based on fixed domicile, i.e. each agent has a fixed housing 
unit. Under unrestricted conditions, or no lockdown, the agents are free to move over the full domain 
using a simple motion model. These motions are independent across agents and encourage smooth 
paths. Under lockdown conditions, most of the agents head directly 
to their housing units and generally stay there 
during that period. A very small fraction of agents are allowed to move freely under the restrictions. 

The rest of this paper is organized as follows. Section 2 develops the proposed agent-level pandemic simulation 
(ALPS) model, specifying the underlying assumptions and motivating model choices. It also discusses
choices of model parameters and present a validation using comparisons with the SIR model. 
Section 3 presents some illustrative examples and discussed computational complexity of ALPS. 
The use of ALPS in understanding influences of countermeasures is presented in Section 4. The paper 
ends by discussing model limitations and suggesting some future directions. 

\section{Agent-Level Pandemic Simulation (ALPS) Model}
In this section we develop our 
simulation model for agent-level interactions and spread of the infections
across a population in a well-defined geographical domain. 
In terms of the model design, there are competing requirements for 
such a simulation to be useful. Our main considerations in the design of 
ALPS are as follows. On one hand, we want to capture detailed properties of agents and
their pertinent environments so as to render a realistic model of pandemic evolution with or without 
countermeasures. On the other hand, we want to keep model complexity reasonably low, in order
to utilize it for analysis under variable conditions and countermeasures. Also, 
to obtain statistical summaries of pandemic conditions under different scenarios, we want to run a large
number of simulations and compute averages.  This also requires keeping the overall model tractable 
from a computational perspective to allow for multiple runs of ALPS in short time. 

\subsection{Simplifying Assumptions}
The overall setting of the simulation model is the following. 
We assuming that the community is based in a square geographical region $D$ 
with $h$ household units arranged in a uniformly-spaced square grid.
We assume that there are $N$ total agents in the community and the configuration updates
every unit interval (hour) counted by the variable $t$. The agents are fully mobile to traverse all of $D$ 
when unconstrained but are largely restricted to their home units under restrictions. 

Next, we specify the assumptions/models being 
used at the moment.
\begin{itemize}
\item {\bf Independent Agents}: 
Each agent (person) has an independent motion model and independent infection probability. The actual infection 
event is of course dependent on being in a close proximity of an infected carrier 
(within a certain distance, say $\approx$ 6 feet) for a certain exposure time. But the probability of 
an agent being infected
is independent of such events for other persons. 

\item {\bf Full Mobility In Absence of Restrictive Measures}: We assume that each agent is fully mobile and moves
across the domain freely when no restrictive measures are imposed. In other words, there no agents with 
restricted mobility due to age or health. 
Also, we do not impose any day/night schedules on the motions, we simply follow a fixed motion 
model at all times. Some papers, including \cite{dunham-spatial-2005}, provide a two- or three-state models 
where the agents transition between some stable states (home, workplace, shopping, etc) in a predetermined
manner. 

\item {\bf Homestay During Restrictive Measures}: We assume that most agents stay at home at all times during 
the restrictive conditions. Only a small percentage (set as a parameter $\rho_0$) of 
population is allowed to move freely
but the large majority stays at home. 

\item {\bf Sealed Region Boundaries}: In order to avoid complications of introducing a transportation model in the system, 
we assume that there is no transfer of agents into and out of the region $D$. The region 
is modeled to have reflecting boundaries
to ensure that all the citizens stay with in the region. The only way the population of $D$ is changed is through 
death of agent(s). 

\item {\bf Fixed Domicile}: The whole community is divided into a certain number of 
living units (households/buildings). These units are placed in square
blocks with uniform spacing. 
Each agent has a fixed domicile at one of the units. During a lockdown period, 
the agents proceed to and stay at home 
with a certain fidelity. We assume that all agents within a unit are exposed to each other, i.e. they are
in close proximity, and can potentially infect others. 

\item {\bf No Re-Infection}: We assume that once a person has recovered from the diseased then he/she can not 
be infected again for the remaining observation period. 
While this is an important unresolved issue for the current Covid-19 infections, it has been a valid assumption 
for the past Corona virus infections. 

\item {\bf Single Patient Ground Zero}: The infection is introduced in the population using a {\bf single carrier}, 
termed patient ground zero at time $t=0$. This patient is selected randomly from the 
population and the time variable is indicated related to this introduction event. 

\item {\bf Constant Immunity Level}: The probability of infection of agents, under the exposure conditions, 
remains same over time. We do not assume any increase or decrease in agent immunity over time. 
Also, we do not assign any age or ethnicity to the agents and all agents are assumed to have equal immunity levels. 

\end{itemize}

\subsection{Model Specifications}
There are several parts of the model that require individual specifications. These parts
include a model on dynamics of individual agents  (with or without restrictions in place), 
the mechanisms of transmitting infections from agent to agents, 
and the process of recovery and 
fatality for infected agents.  A full listing of the model parameters and some typical 
values are given in Table 1 in the appendix. 

\begin{itemize}
\item {\bf Motion Model}: The movement of a subject follows a simple model where the 
instantaneous velocity $v_i(t)$ is a weighted sum of three components: (1) 
velocity at the previous time, \ie $v_i(t-1)$, (2) a directed component guiding them to their home, 
$\alpha (h_i - x_i(t-1))$, 
and (3) an independent Gaussian increment $\sigma w_i(t)$, $w_i(t) \sim {\cal N}(0,1)$. 
Note that the motions of different agents are kept independent of each other. 
The located $h_i \in D$ denotes the home unit (or stable state) of the $i^{th}$ person. 

Using mathematical notation, the model for instantaneous position $x_i(t)$ 
and velocity $v_i(t)$ of the $i^{th}$ 
agent are given by: 
\begin{eqnarray*}
v_i(t) &=& \mu~ v_i(t-1)  +  (1 - \mu)~\alpha (h_t - x_i(t-1)) + \sigma w_i(t),\ 0 \leq \mu \leq 1\ , \\
x_i(t) &=& x_i(t-1) + \delta v_i(t) \ .
\end{eqnarray*}
Here $\alpha \in \real_+$ determines how fast one moves towards their home, and $\mu$ quantifies 
the degree to which one follows the directive to stay home. If $\mu = 0$, then the person reaches home 
and stays there except for the random component $w_i$. However, if $\mu = 0.5$, then a significant 
fraction of motion represents continuity irrespective of the home location. The value $\mu = 1$
implies that  
either there is not restriction in place or the person does not follow the directive.

\noindent {\bf Reflecting Boundary}: When a subject reaches boundary of the domain $D$, the 
motion is reflected and the motion continues in the opposite direction. 

Fig.~\ref{fig:examp-motion} shows examples of random agent motions under different 
simulation conditions. The leftmost case shows when there is no lockdown and the agents are moving 
freely throughout. The middle case shows the case when the restrictions are imposed on day 10 and the restrictions 
stay in place after that. The last plot shows the case where a lockdown is imposed on day 10 
and then lifted on day 20. 
\begin{figure}[h]
\begin{center}
\begin{tabular}{ccc}
\includegraphics[height=1.5in]{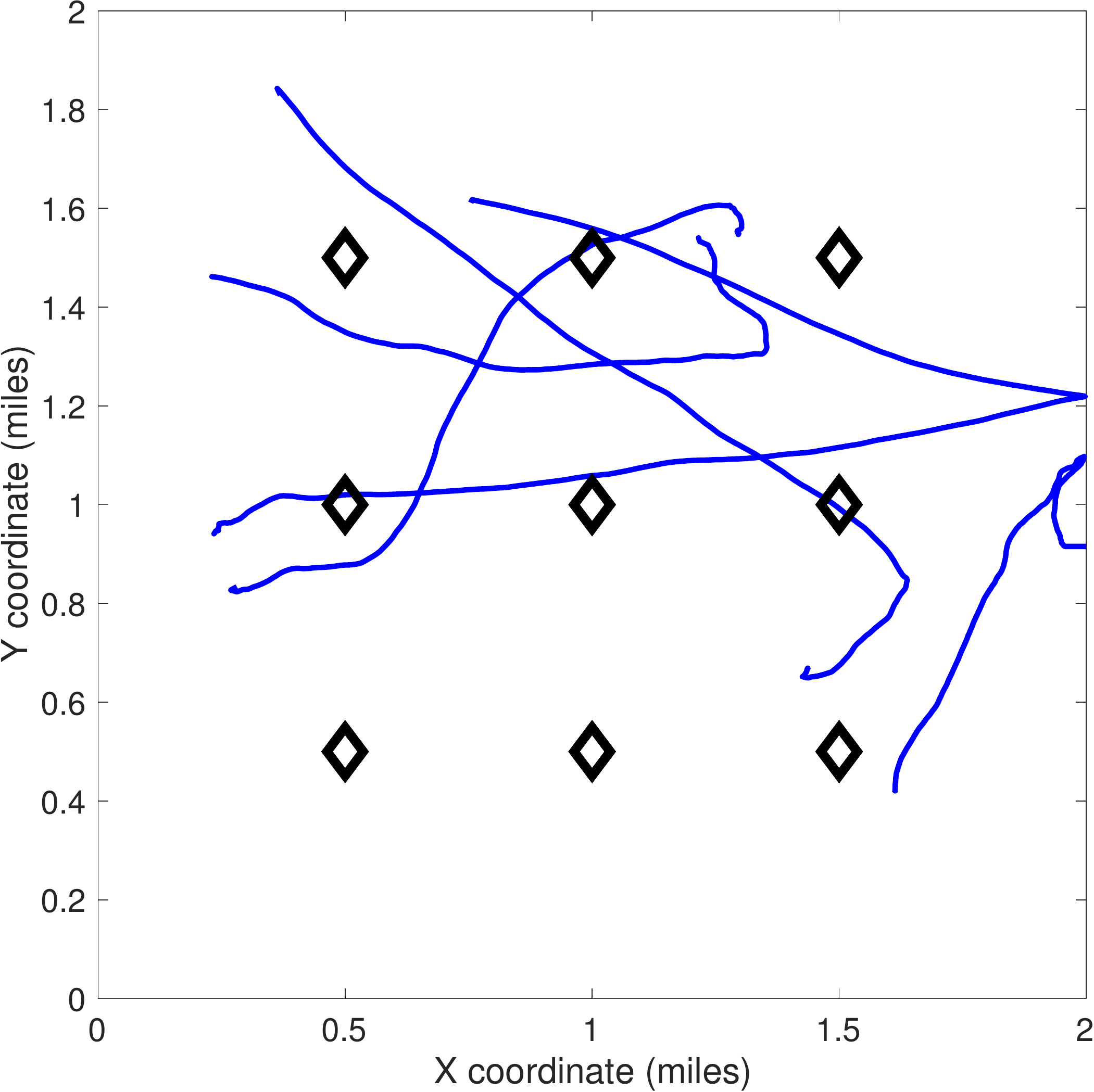} &
\includegraphics[height=1.5in]{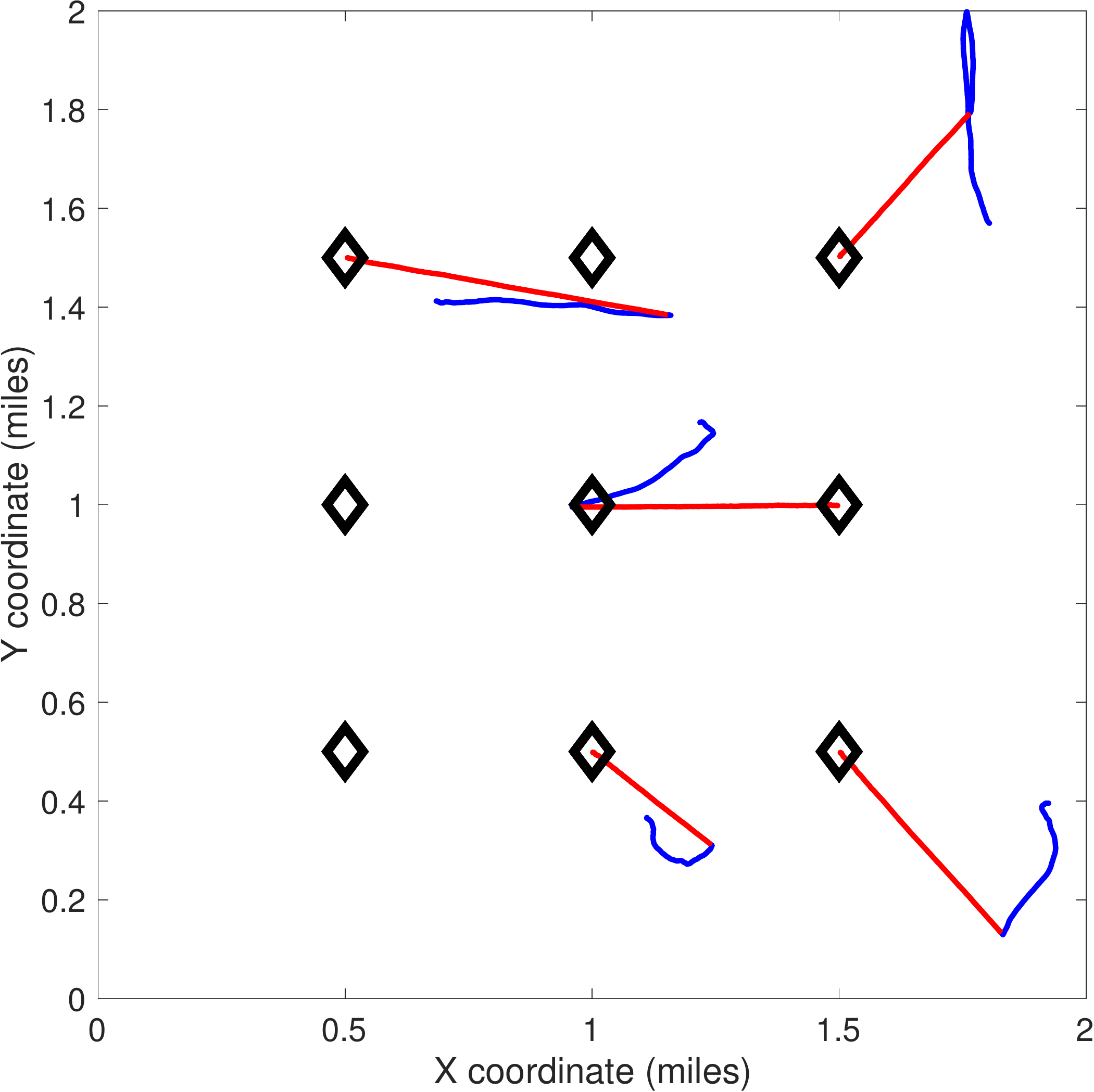} &
\includegraphics[height=1.5in]{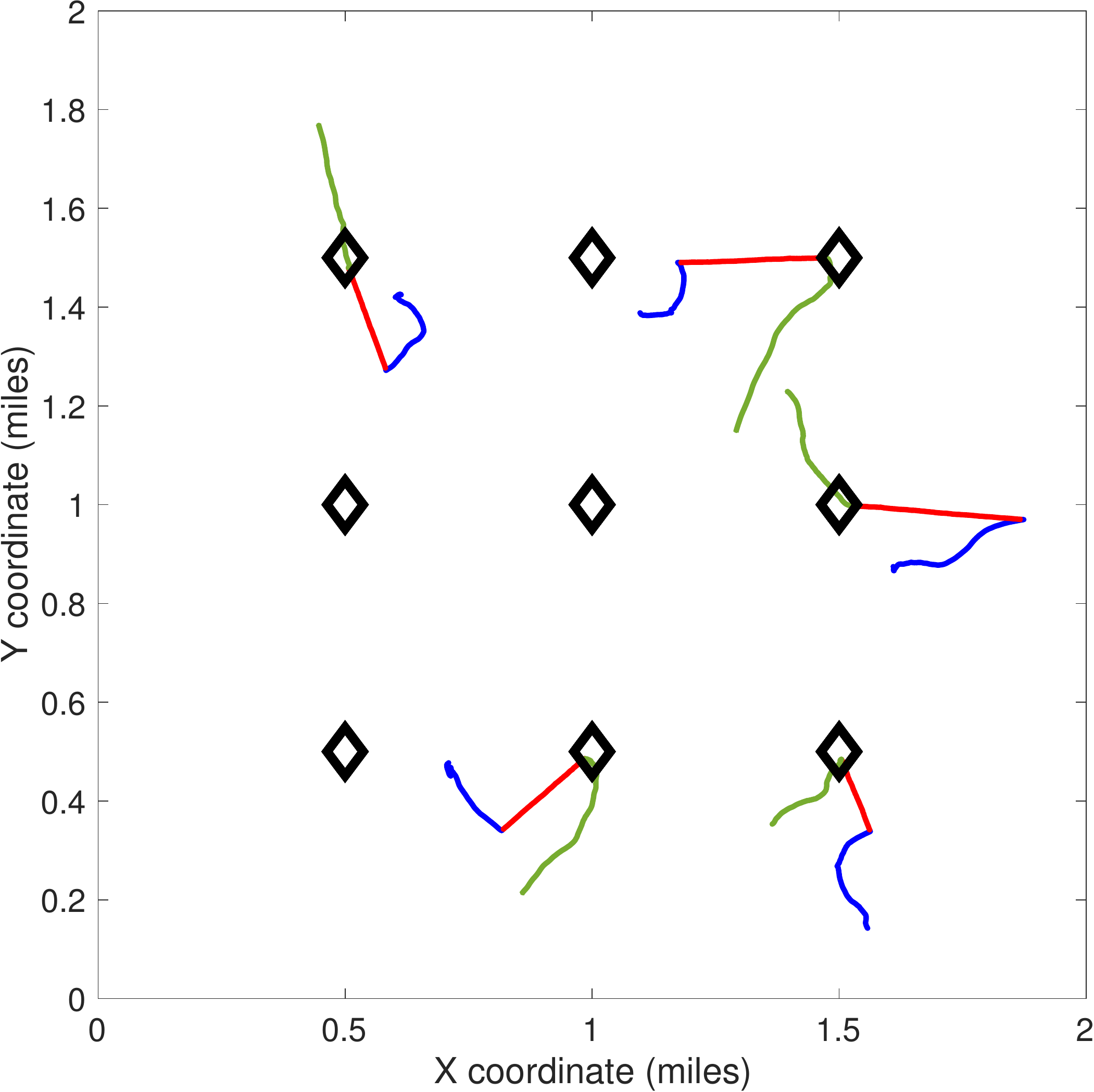} \\
No Lockdown & Lockdown at Day 10 & Lockdown during Day 10-20
\end{tabular}
\caption{Motions of five randomly selected agents under different conditions.
Blue curves denote unrestricted movements, red curves denote movement during lockdown, and green
curves denote movement after lockdown.} \label{fig:examp-motion}
\end{center}
\end{figure}

\item {\bf Restrictions (or Lockdown) Model}: 
Once the restriction period starts, at time $T_0$, all agents are directed towards their homes 
and asked to stay there. We assume that $\rho \%$ of the subject 
follow this directive while the others ($(100-\rho)\%$) follow a different motion model. 
The variable $\rho$ changes over time according to:
\begin{equation}
\rho(t) = \left\{ \begin{array}{ll}  0, & t < T_0 \ \mbox{(under no restrictions)}\\
 \rho_0, & t \geq T_0 \ \mbox{(under restrictions)\ .} \end{array}
\right.
\end{equation}
We note that the people who do not follow restrictions  follow the prescribed motion 
model with $\mu= 1$.

\item {\bf Exposure-Infection Model}: 
The event of infection of an agent depends on the level of exposure to another infected person. 
This process is controlled by the following parameters:
\begin{itemize}
\item The distance between the subject and the 
infected person should be less than $r_0$.
\item The amount of exposure in terms of the number of time units should be at least $\tau_0$. At 
the moment we use the cumulative exposure over the whole history, rather than just the recent history. 
\item Given that these conditions are satisfied, the probability of catching the disease at each time $t$ is an 
independent Bernoulli random variable with probability $p_I$. 
\end{itemize}

\item {\bf Recovery-Death Model}: Once a subject is infected, we randomly associate the nature 
of the illness immediately. Either an infected agent is going to recover (non-fatal type, or NFT) or the person is 
eventually going to die (fatal type, or FT). The probability of having a fatal infection given that a person 
is infection is $p_F$. 
\begin{itemize}
\item {\bf Recovery}: A subject with a non-fatal type (NFT) is sick for a period of $T_R$ days. After 
this period, the person can recover at any time according a Bernoulli random variable with probability $p_R$. 

\item {\bf Fatality}: A subject with a fatal type (FT) is sick for a period of $T_D$ days. After 
this period, the person can die at any time 
independently according a Bernoulli random variable with probability $p_D$. 

\end{itemize}

\end{itemize}

\subsection{Chosen Parameter Values }

A complete listing of the ALPS model parameters is provided in the Appendix in 
Table 1. In this section, we motivate the values chosen for those
parameters in these simulations. These values are motivated by the current reports
of Corona virus pandemic.

\begin{itemize}

\item We have used a square domain with size 2 miles $\times$ 2 miles 
for a community with population of 
N agents. For $N=900$, the model represents a 
population density of 225 people/mile$^2$.  The community contains
$h$ living units (buildings) with a domicile of $N/h$ people per unit. 
In case $N/h$ is high, a unit represents a tall building in metropolitan areas, but when 
$N/h$ is small, a unit represents a single family home in a suburban area. 

The time unit for updating configurations is one hour and occurrence of major events is specified 
in days. For example, the lockdown can start on day 1 and end on day 60.

\item The standard deviation for accelerations in agent mobility are approximately 1-5 feet/hour (fph). 
Through integration over time,
this results in agent speeds up to 1000 fph. We assume that $\rho_0 = 0.98$, \ie 98\% of the people follow
the restriction directives. 

\item The physical distance between agents to catch infection should at
most $r_0 \approx 6ft$
and the exposure time should be at least $\tau_0 =  5$ time units (hours). The probability $p_I$ of 
getting infected, under the right exposure conditions, is set at 5\% at each time unit (hour) independently. 

\item 
Once infected, the probability of having a fatal outcome is set at 5-10\%. 
The time period of recovery for agents with non-fatal outcomes starts at 7 days. The probability 
of reaching full recovery for those agents is $p_R  = 0.001$ at each subsequence time unit (hour). 
Similarly, for the agents with fatal outcomes, the period of 
being infected is set to be 7 days and after that the probability of death at each time unit (hour) is 
set to be $p_D = 0.1$. 
\end{itemize}

\subsection{Model Validation}

Although ALPS is perhaps too simple model to capture intricate dynamics of an active society, 
it does provide an efficient tool for analyzing effects of countermeasures during the
spread of a pandemic. Before it can be used in practice, 
there is an important need to validate it in some way. 

As described in \cite{hunter2017}, there are several ways to validate a
simulation model. One is to
use real data (an observed census of infections over time) in a community to estimate model 
parameters, followed by a statistical model testing. While such data may emerge for Covid-19 for
public use in the future (especially with the advent of tracking apps being deployed in 
many countries), there is currently no such agent-level data available for  
Covid-19.  The other approach for validation is to consider coarse population-level 
variables and their dynamics, 
and compare them against established models such as SIR and its variations. 
We take this latter approach.

\begin{figure}
\begin{center}
\begin{tabular}{cc}
\includegraphics[height=1.5in]{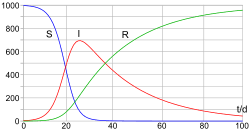}&
\includegraphics[height=1.5in]{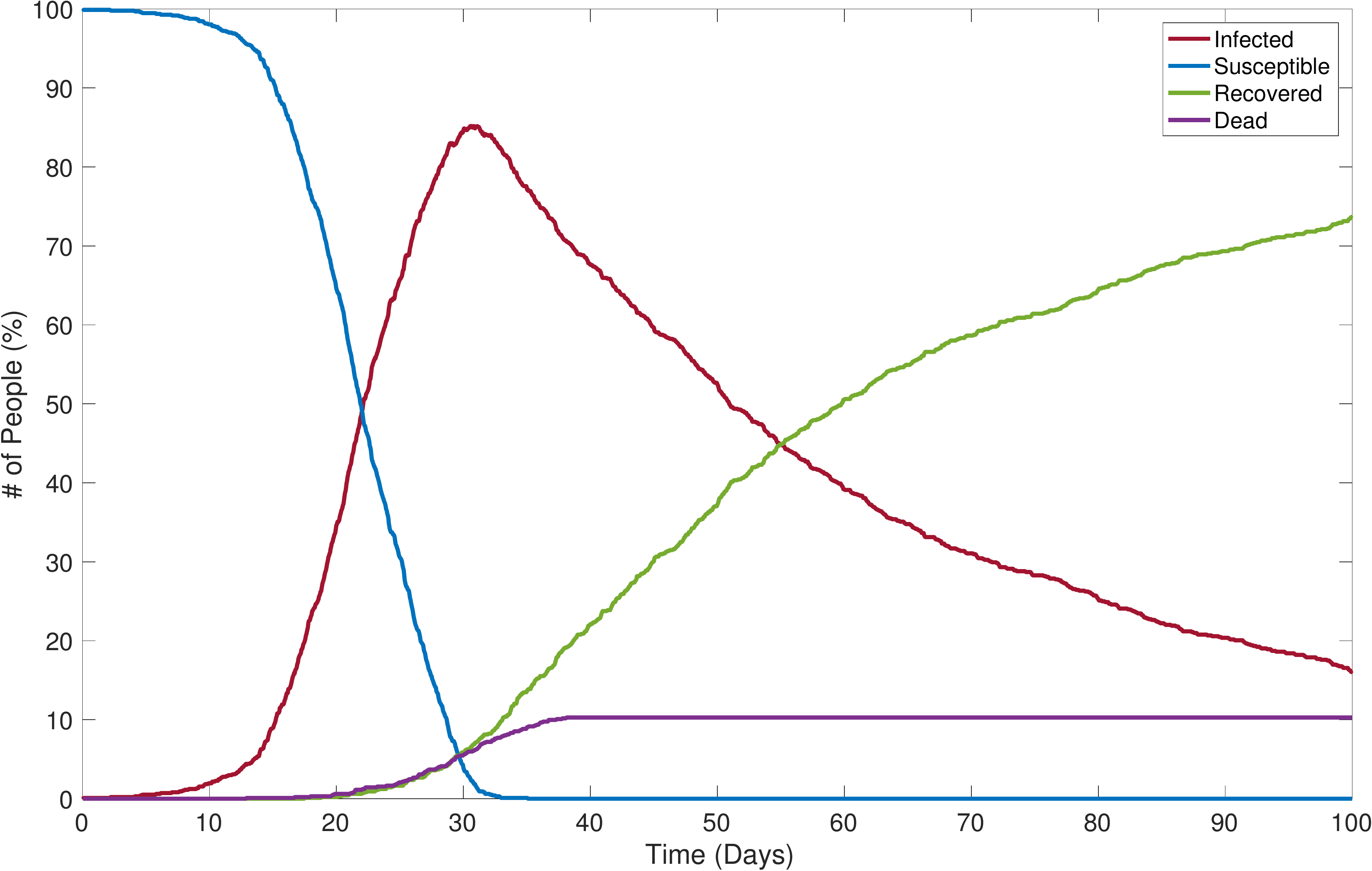}
\end{tabular}
\caption{The evolution of population-level infection measurements under the SIR model (left), 
courtesy wikipedia) and the ALPS model (right).}
\label{fig:validation} 
\end{center}
\end{figure}

Fig.~\ref{fig:validation}  shows plots of the evolutions of global infection counts 
(susceptible, infection, recovered) in a community under the well-known SIR model
(on the left) and 
the proposed ALPS model (on the right). In the ALPS model the counts for recovered and 
fatalities are kept separate, while in SIR model these two categories are combined. 
One can see a remarkable similarity in the shapes of corresponding curves and this provides
a certain validation to the ALPS model. In fact, given the dynamical models of agent-level 
mobility and infections, one can in principle derive parameters of the population-level 
differential equations used in the SIR model. We have left that out for future developments. 

\section{Exemplar Scenarios \& Computational Complexity}
We illustrate the use of ALPS model by showing its outcomes 
under a few typical scenarios. Further, we discuss the computational cost of running 
ALPS on a regular laptop computer. 

\subsection{Examples from ALPS}
We start by showing outputs of ALPS under some interesting settings. 
In these examples, we use a relatively small number of agents ($N = 900$)
and household units ($h = 9$), with $T = 100 days$, in order to improve 
visibility of displays. 

\begin{enumerate}

\item {\bf Example 1 -- No Restrictive Measures}: \\
Fig.~\ref{examp1} shows a sequence of temporal snapshots representing thecommunity at different 
times over the observation period. In this example, the population is fully 
mobile over the observation period and no social distancing restrictions are imposed. 
The snapshots show the situations at 10, 20, 50, and 100 days. 
The corresponding time evolution of global count measures is shown in the bottom right panel.
The infection starts to spread
rapidly around the 10th day and reaches a peak infection level of 81\% around day 35.  Then the 
recovery starts and continues until very few infected people are left. In this simulation, the number of 
fatalities is found to be 11\%. 

\begin{figure}[h]
\begin{center}
\begin{tabular}{ccc}
\includegraphics[height=1.5in]{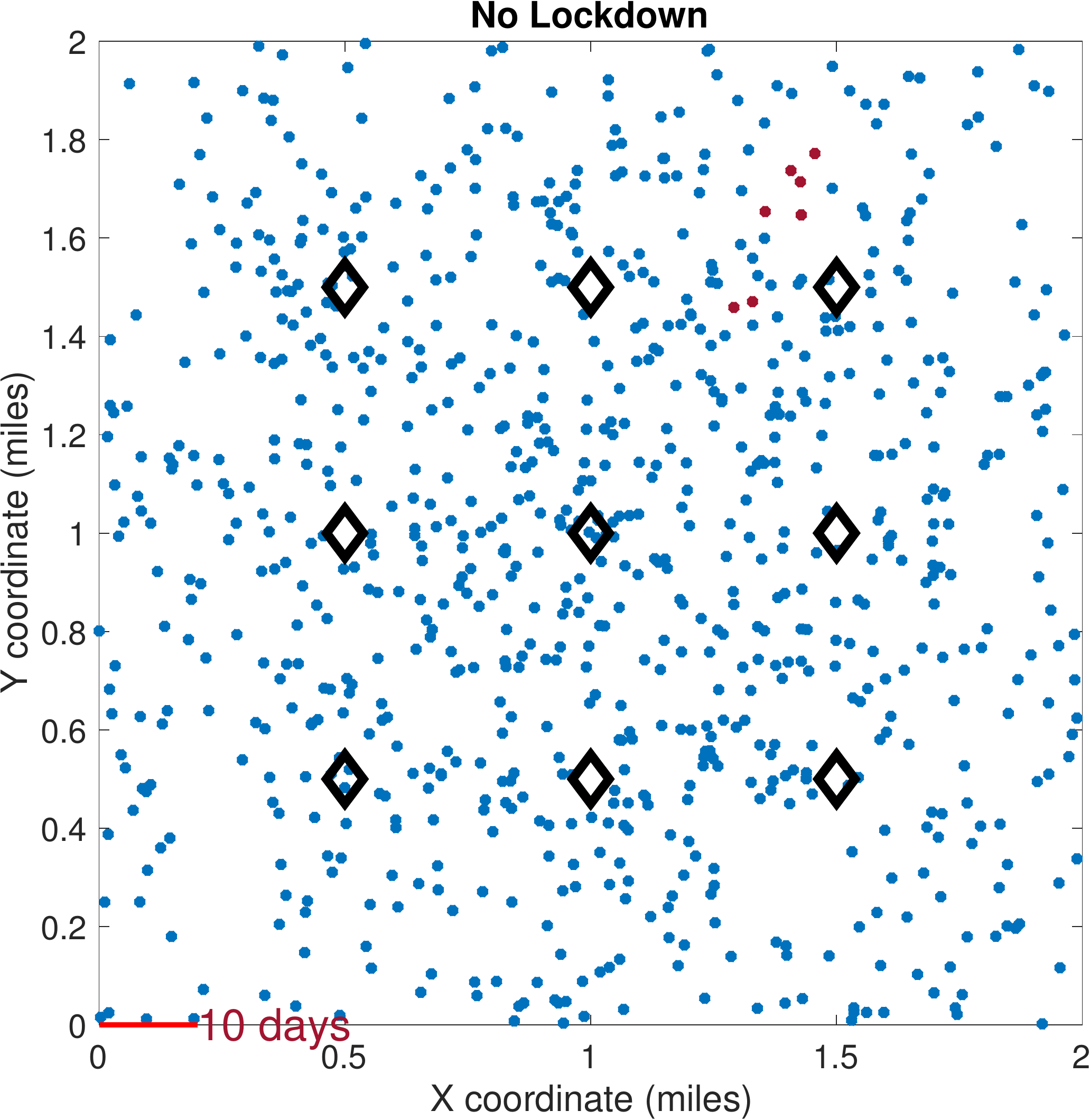}&
\includegraphics[height=1.5in]{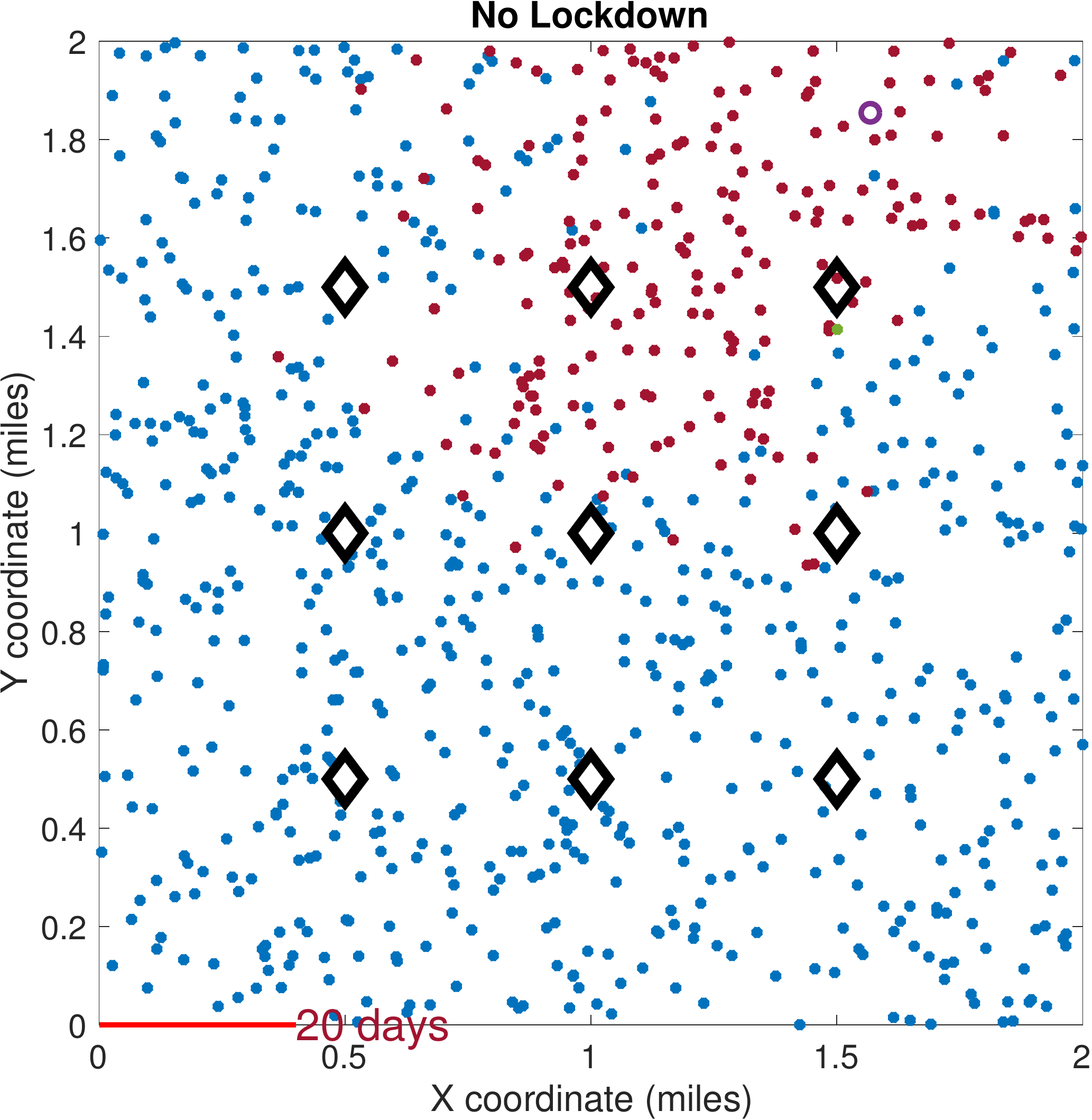}&
\includegraphics[height=1.5in]{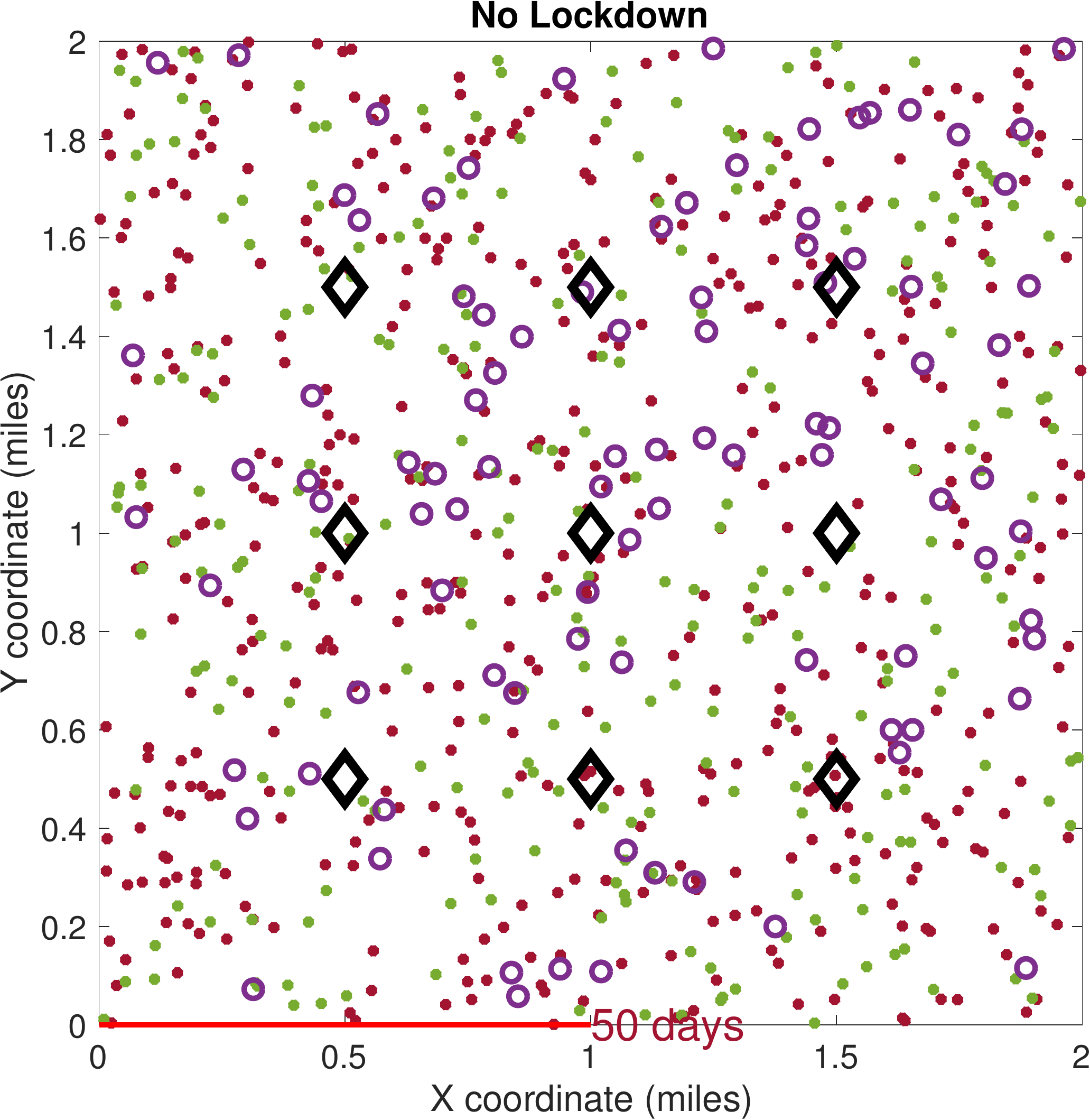}\\
\hspace*{0.2in}\includegraphics[height=1.5in]{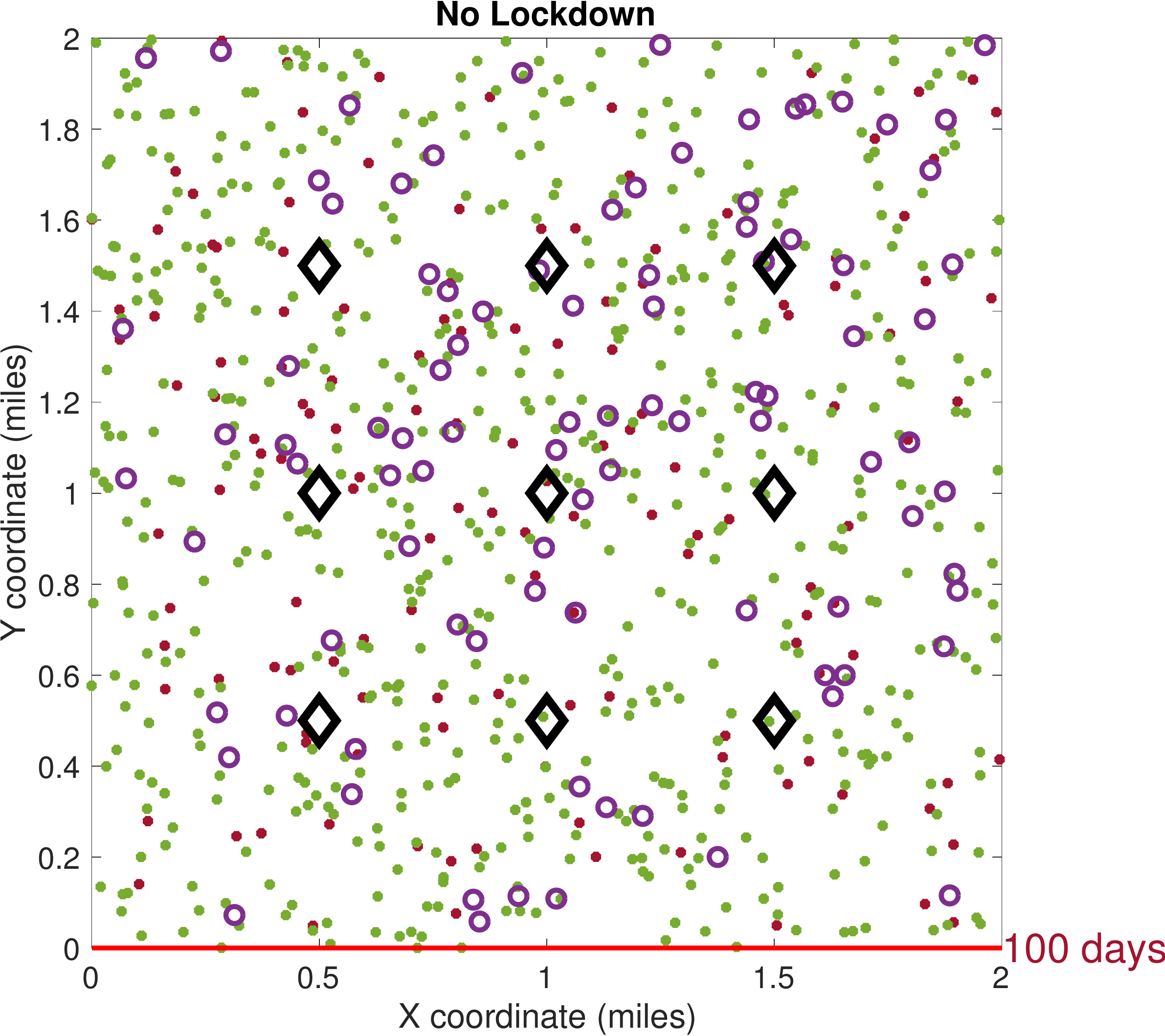} &
&\hspace*{-1.5in} \includegraphics[height=1.5in]{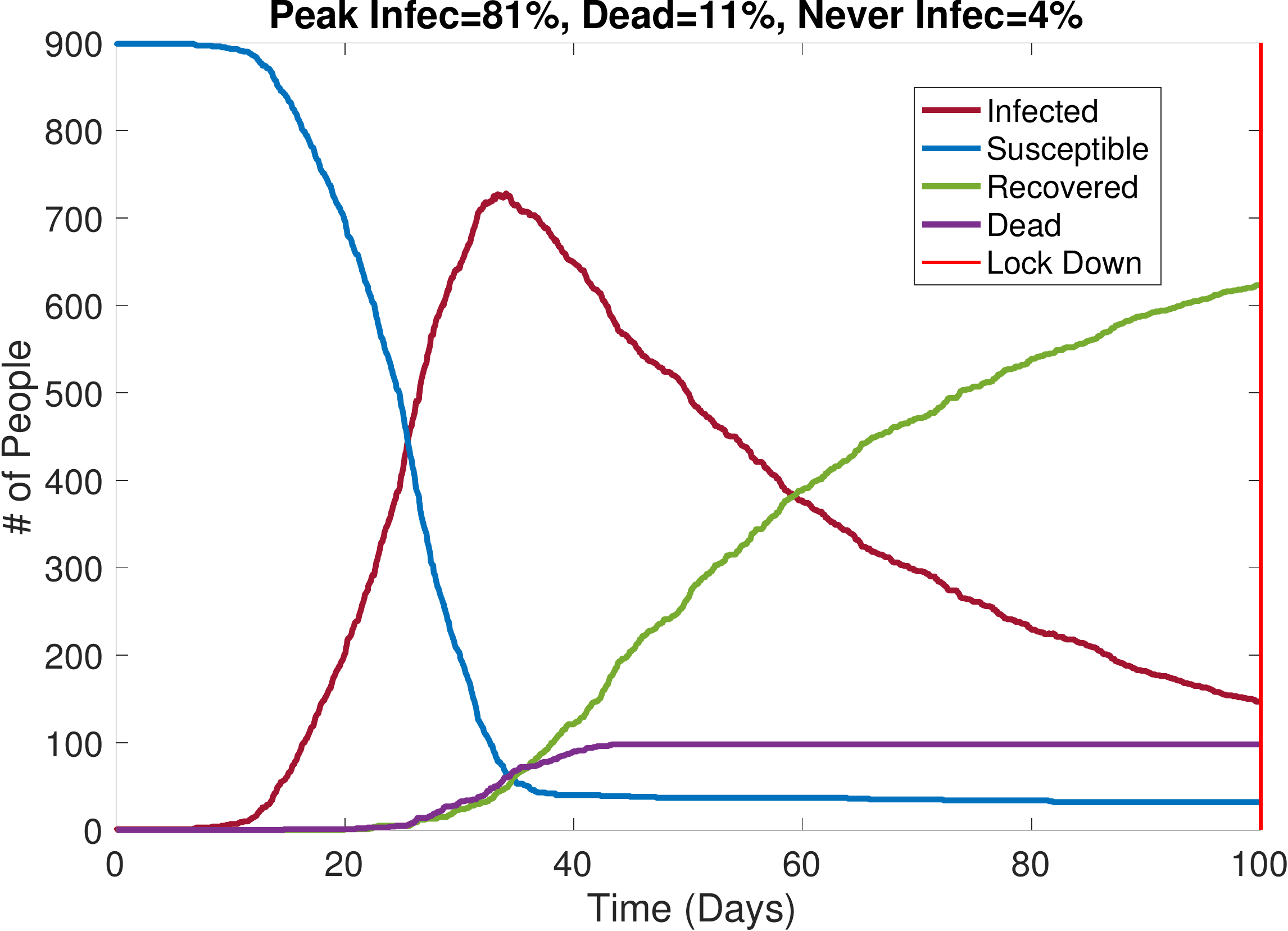}\\
\end{tabular}
\caption{Example 1: Model outputs at different times under no restrictive measures. Blue dots are 
susceptible agents, red dots are infected agents, green dots are recovered agents, and purple circles
denote fatalities.}
\label{examp1}
\end{center}
\end{figure}

\item {\bf Example 2 -- Early Restrictions}: \\
In the second example, the restrictions are introduced on day 5 after the infection and these measures stay in place
after that. The results are shown in Fig.~\ref{examp2}. 
The top left panel shows the situation on day 2 where the population is full mobile and 
the infection has not started spreading yet. The situation on day 6 shows infection beginning to spread
around agent zero and early signs of clustering of agents around their domicile units. 
The scene for day 10 shows progression of lockdown with most agents (98\%) getting close to their 
domicile units and the infection being carried to some units by their resident agents. By day 20, 
the concentration of agents in their units is complete and only a few (2\%)  agents are allowed 
free mobility. 

The bottom right panel shows temporal evolutions of the population-level infection counts: 
susceptible (blue curve), infected (red), recovered (green), and deceased (magenta).  
As the picture shows, the infections start growing initially but the gains of lockdown measures
start appearing around day 15 -- it takes about 10-12 for the restrictions to show results. The subsequent
bumps in the infected counts is due to the new infections transmitted by roaming agents. 
In this run of ALPS, we see an overall fatality rate of 3\% and an uninfected population size of 67\%.

\begin{figure}[h]
\begin{center}
\begin{tabular}{ccc}
\includegraphics[height=1.5in]{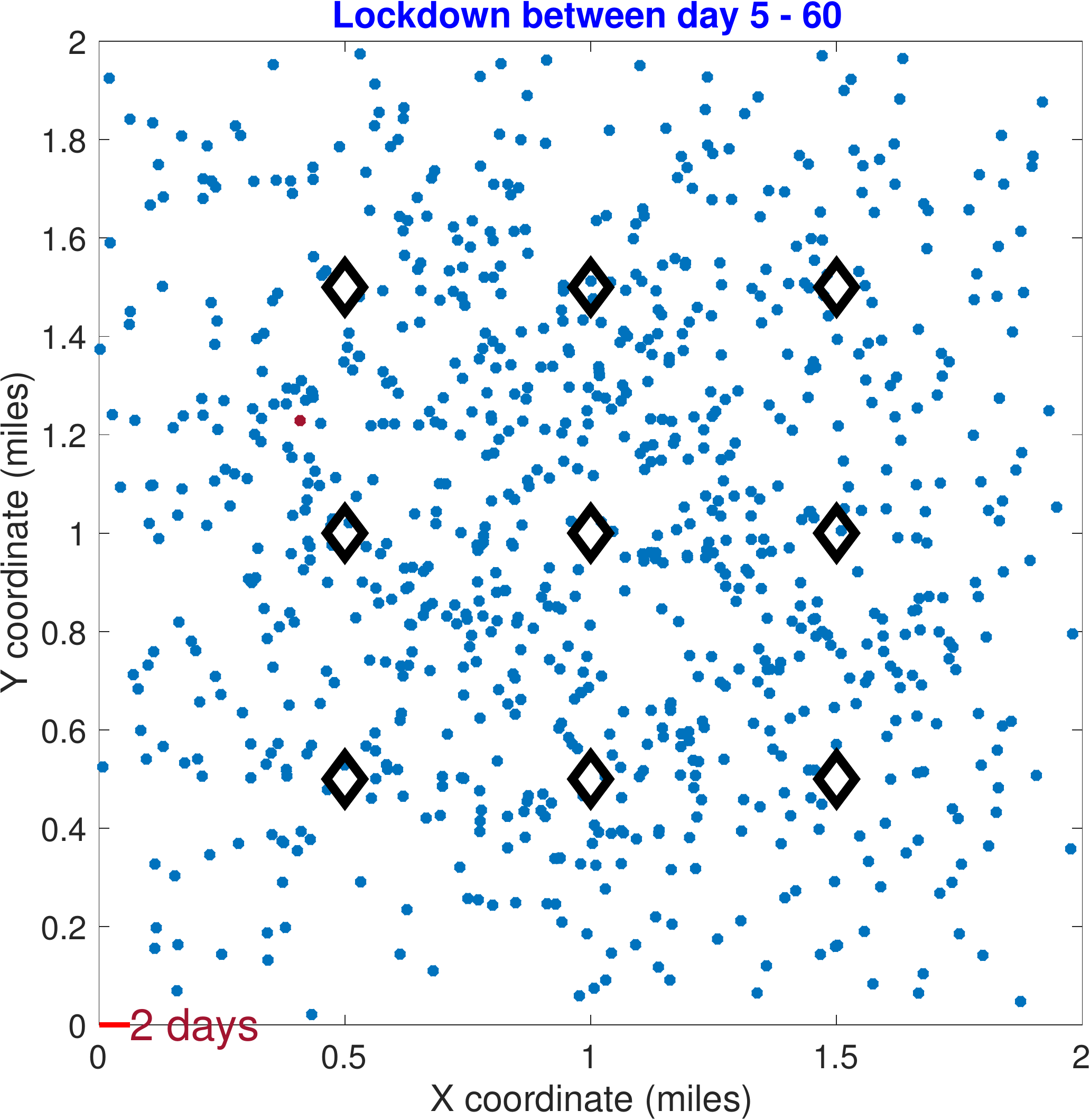}&
\includegraphics[height=1.5in]{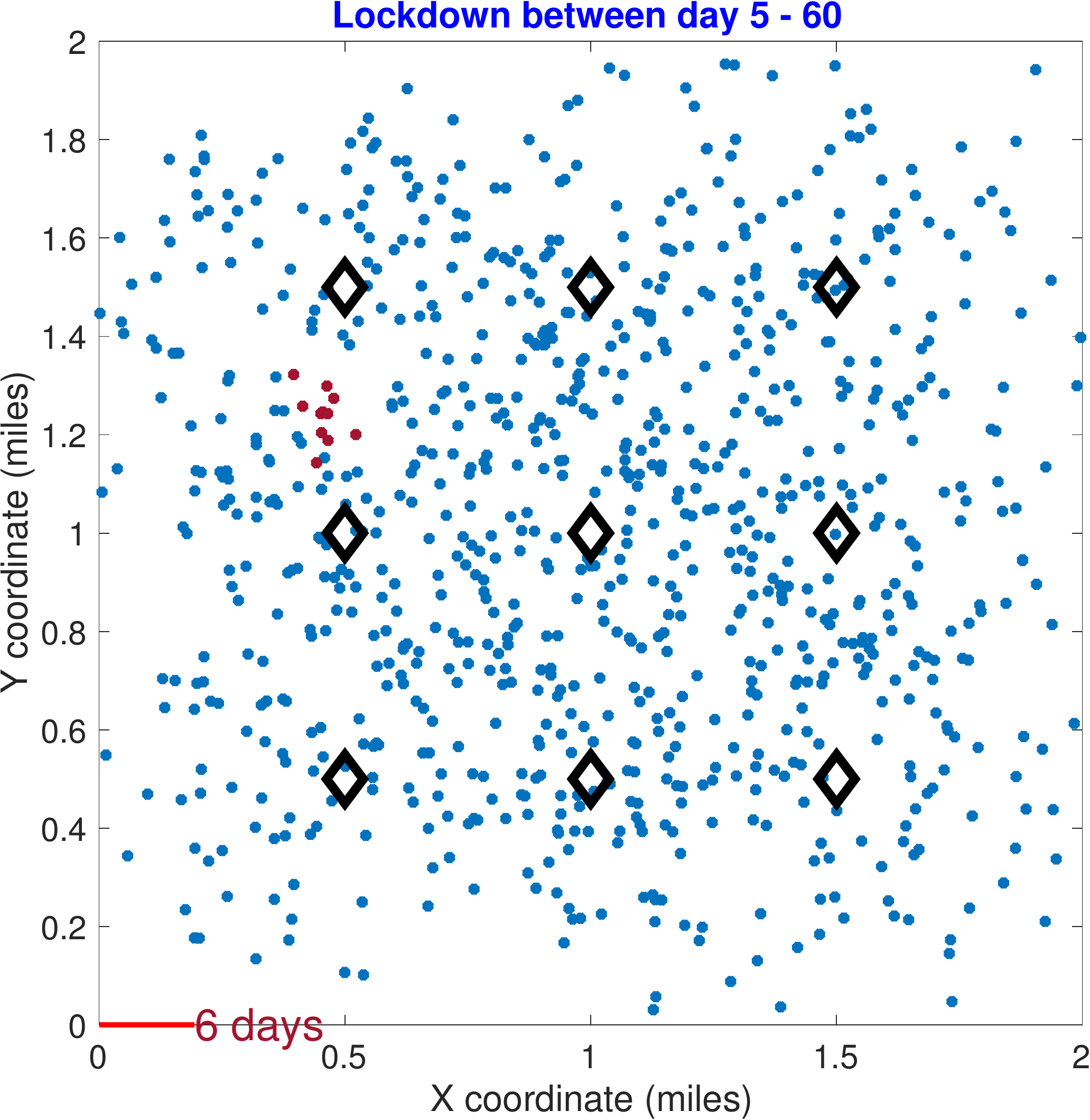}&
\includegraphics[height=1.5in]{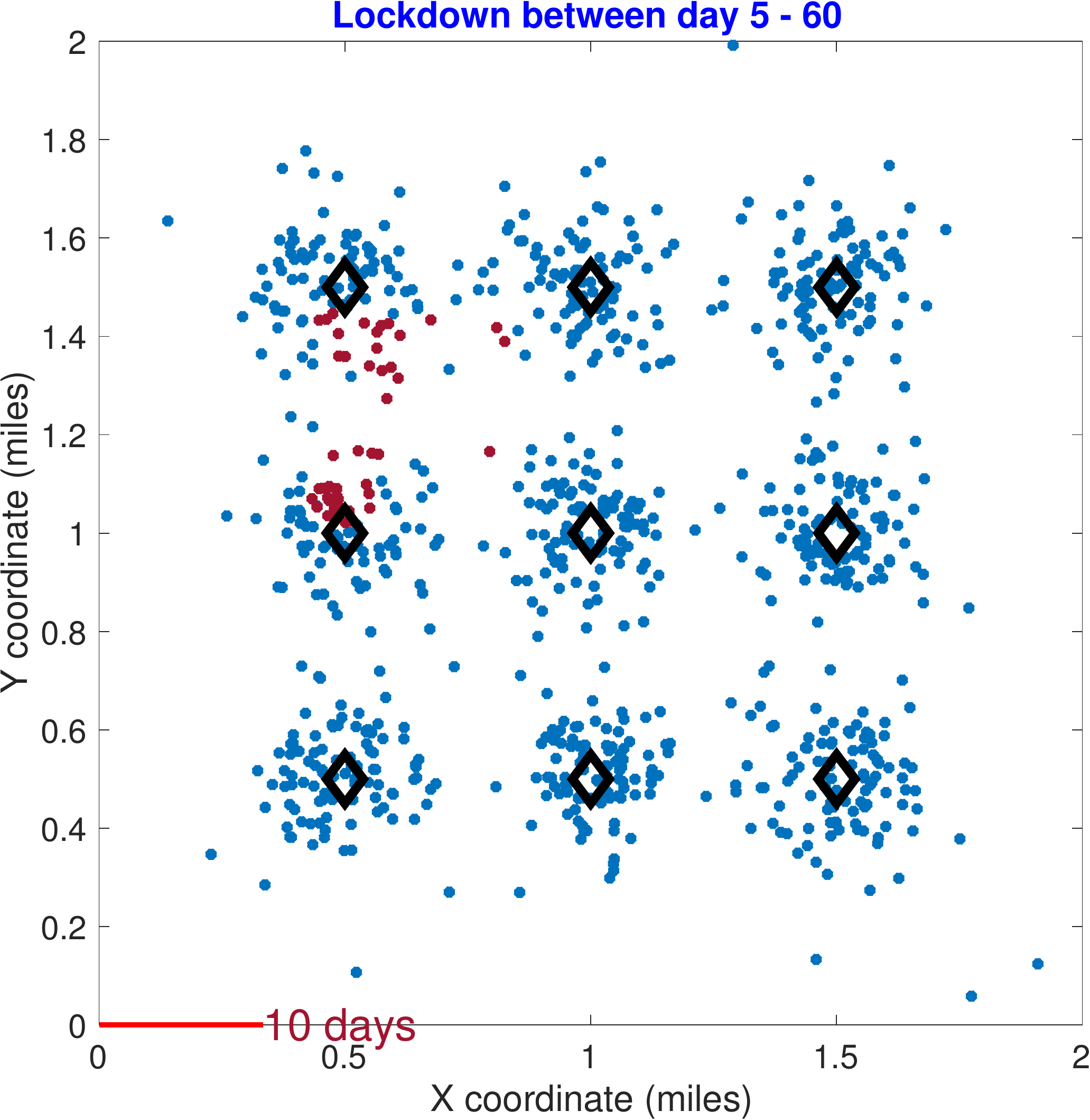}\\
\includegraphics[height=1.5in]{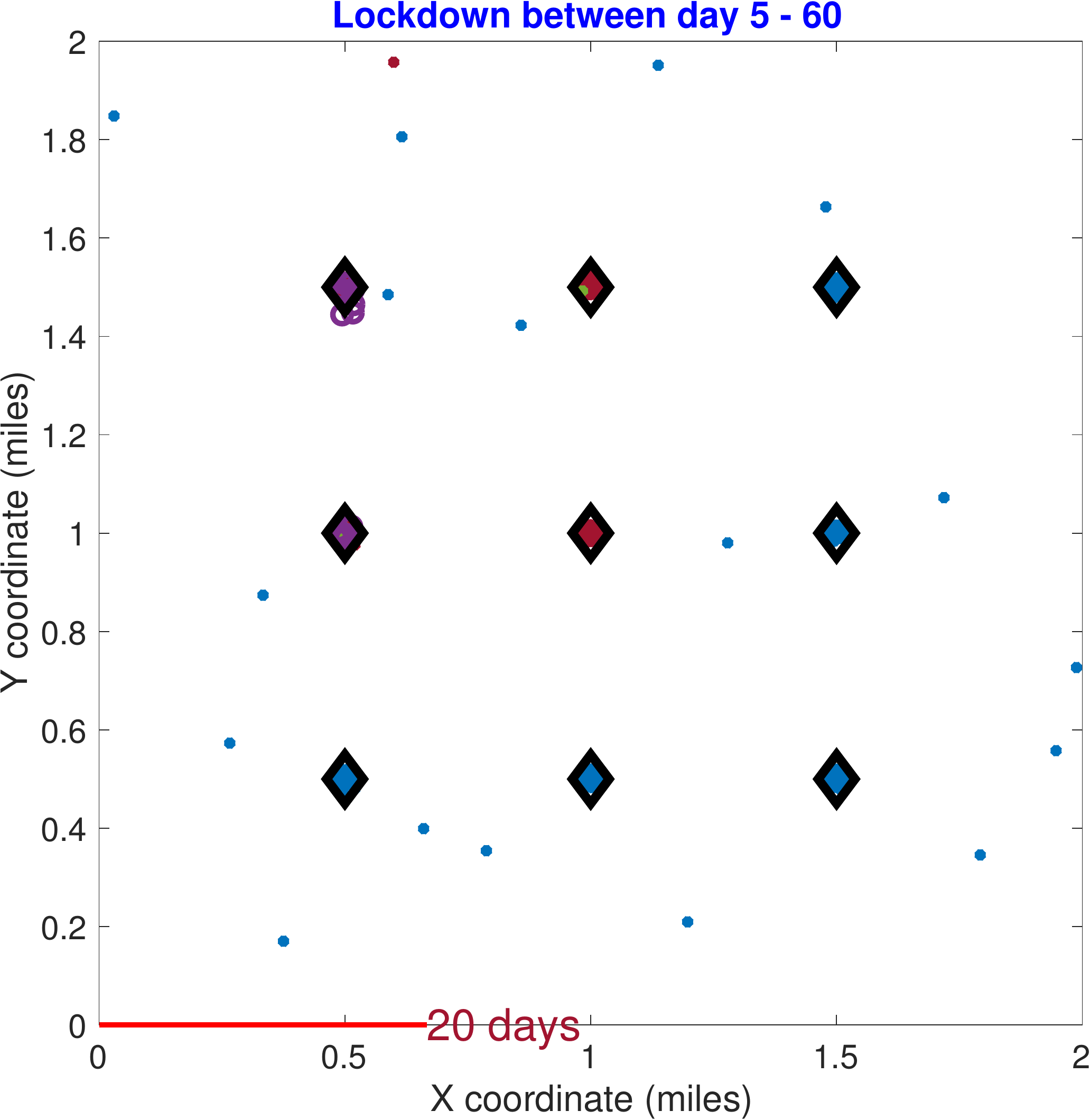} &
&\hspace*{-1.5in} \includegraphics[height=1.5in]{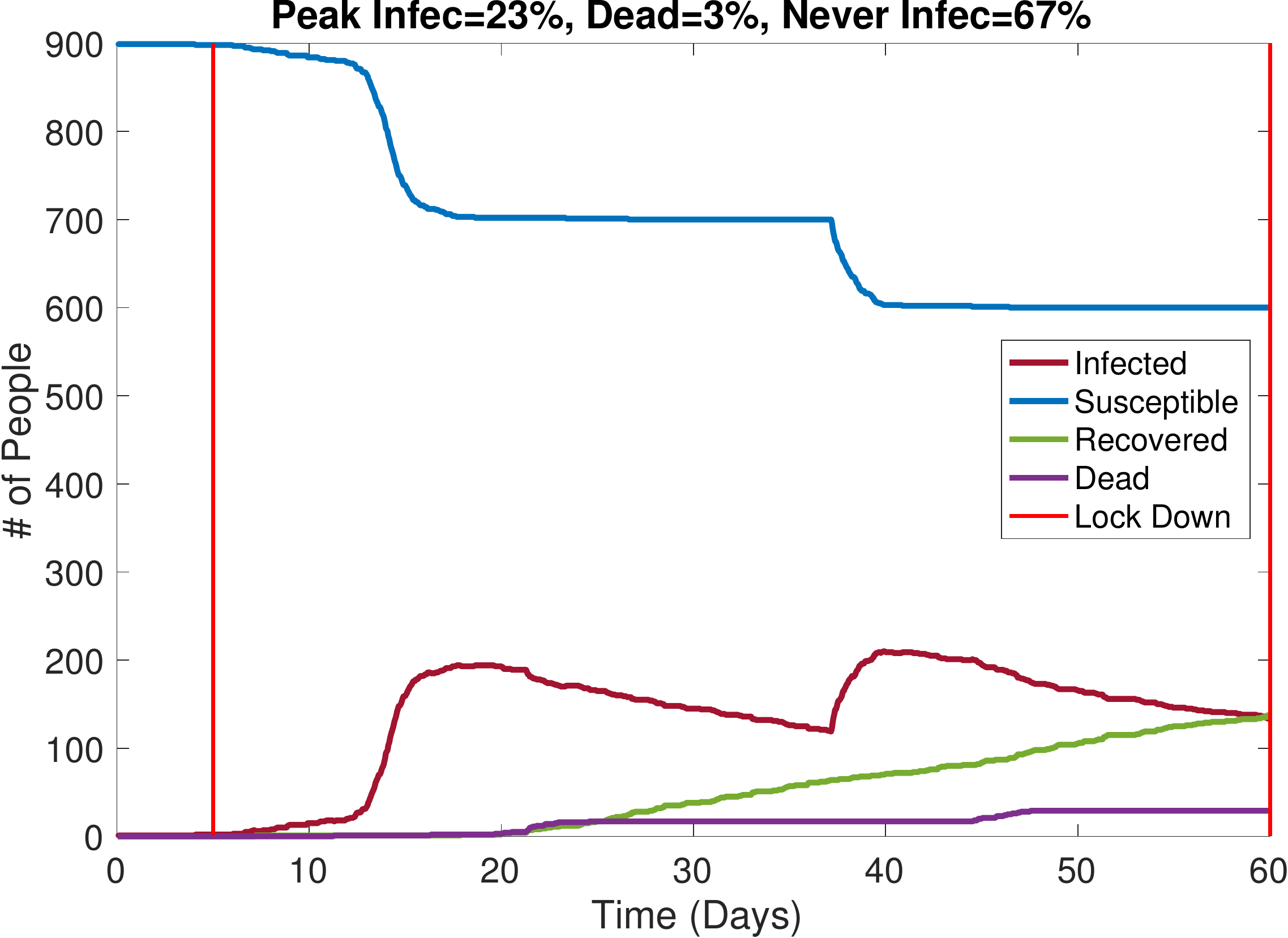}\\
\end{tabular}
\caption{Example 2: Model outputs when the restrictive measures are imposed on day 5.}
\label{examp2}
\end{center}
\end{figure}

\item {\bf Example 3: Early Imposition and Then Removal of Restrictions }:\\
In this example, with results shown in Fig.~\ref{examp3}, 
the restrictions are introduced on day 5 after the infection and 
lifted on day 30. So the restriction period of 25 days is surrounded by unrestricted
mobility on both sides.  As the plot of global variables indicate, the early restriction
helps reduce the infection rates but these gains are lost soon after lifting of the
restrictions. The percentage of infected people goes back up and the rate of fatalities
resemble the unrestricted situation in Example 1. 

\begin{figure}[h]
\begin{center}
\begin{tabular}{ccc}
\includegraphics[height=1.5in]{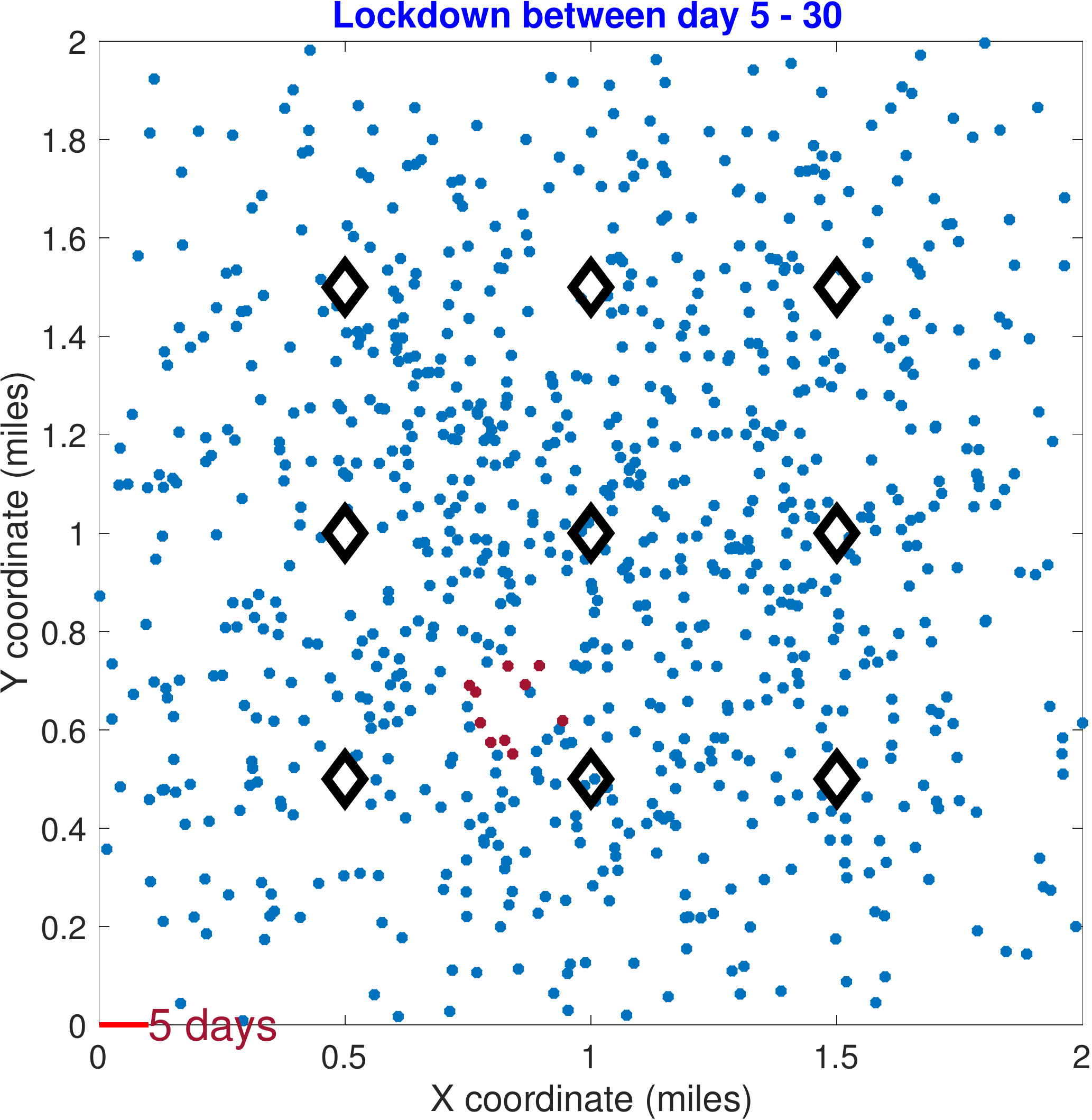}&
\includegraphics[height=1.5in]{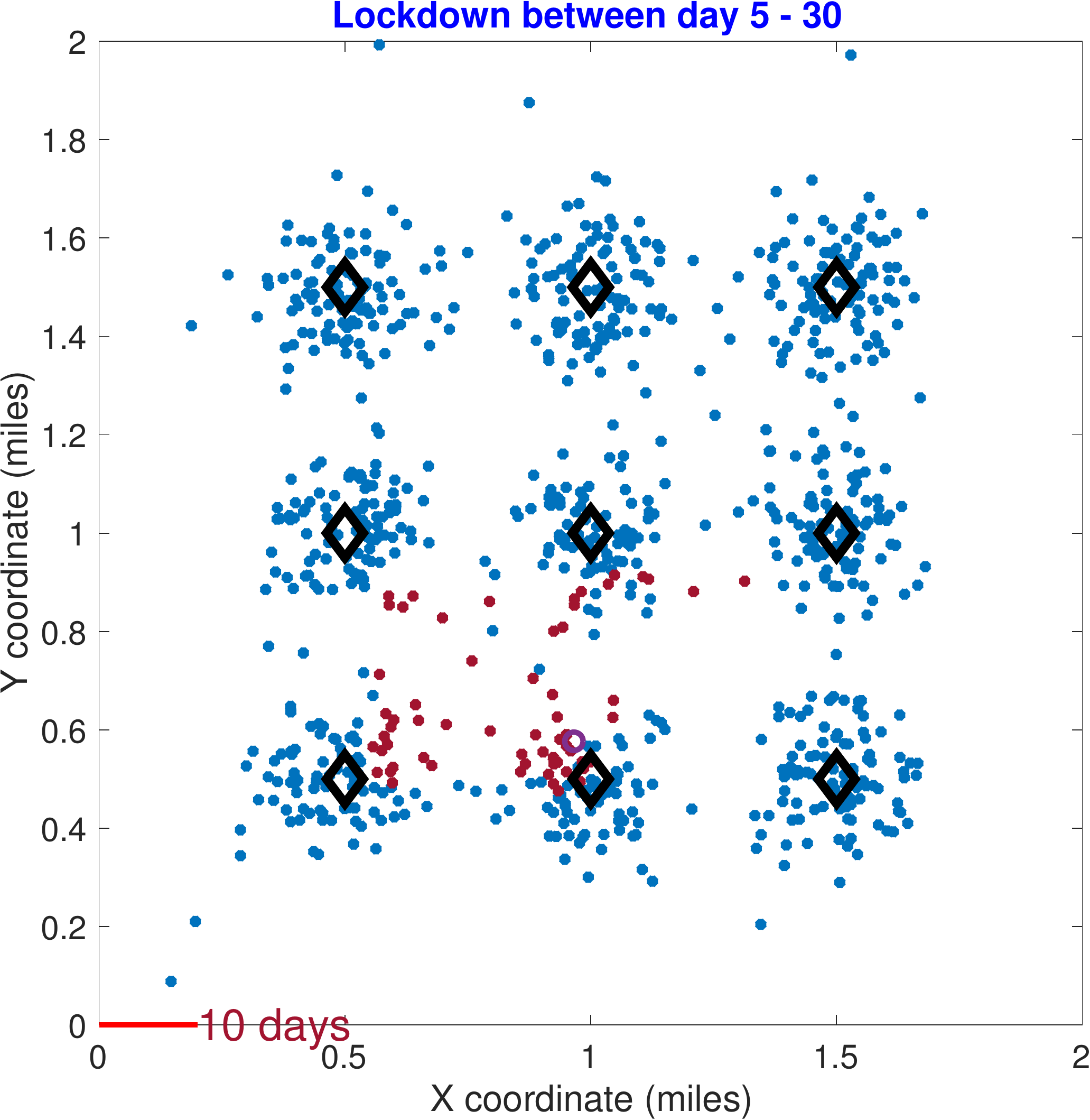}&
\includegraphics[height=1.5in]{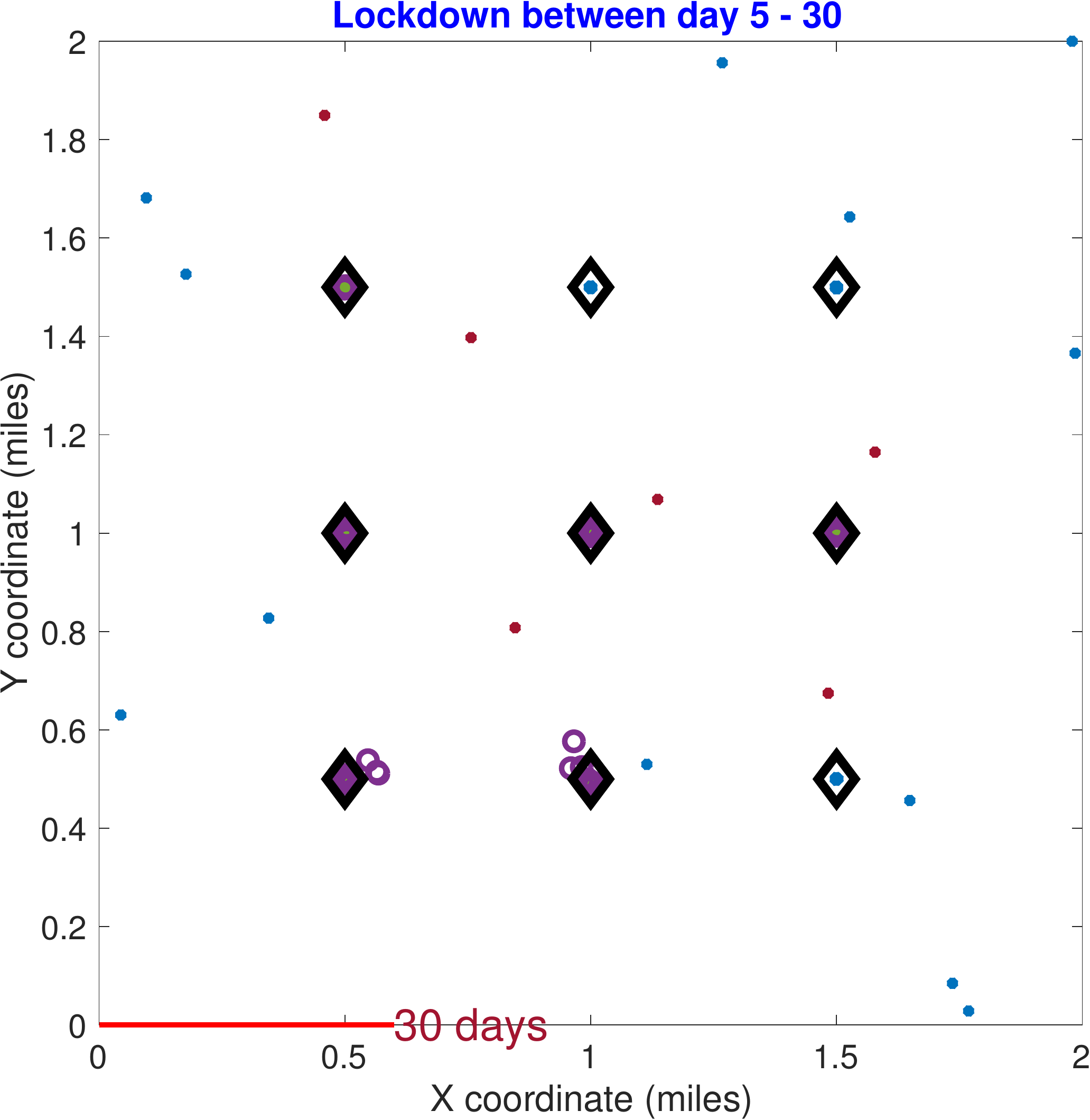}\\
\hspace*{0.2in}\includegraphics[height=1.5in]{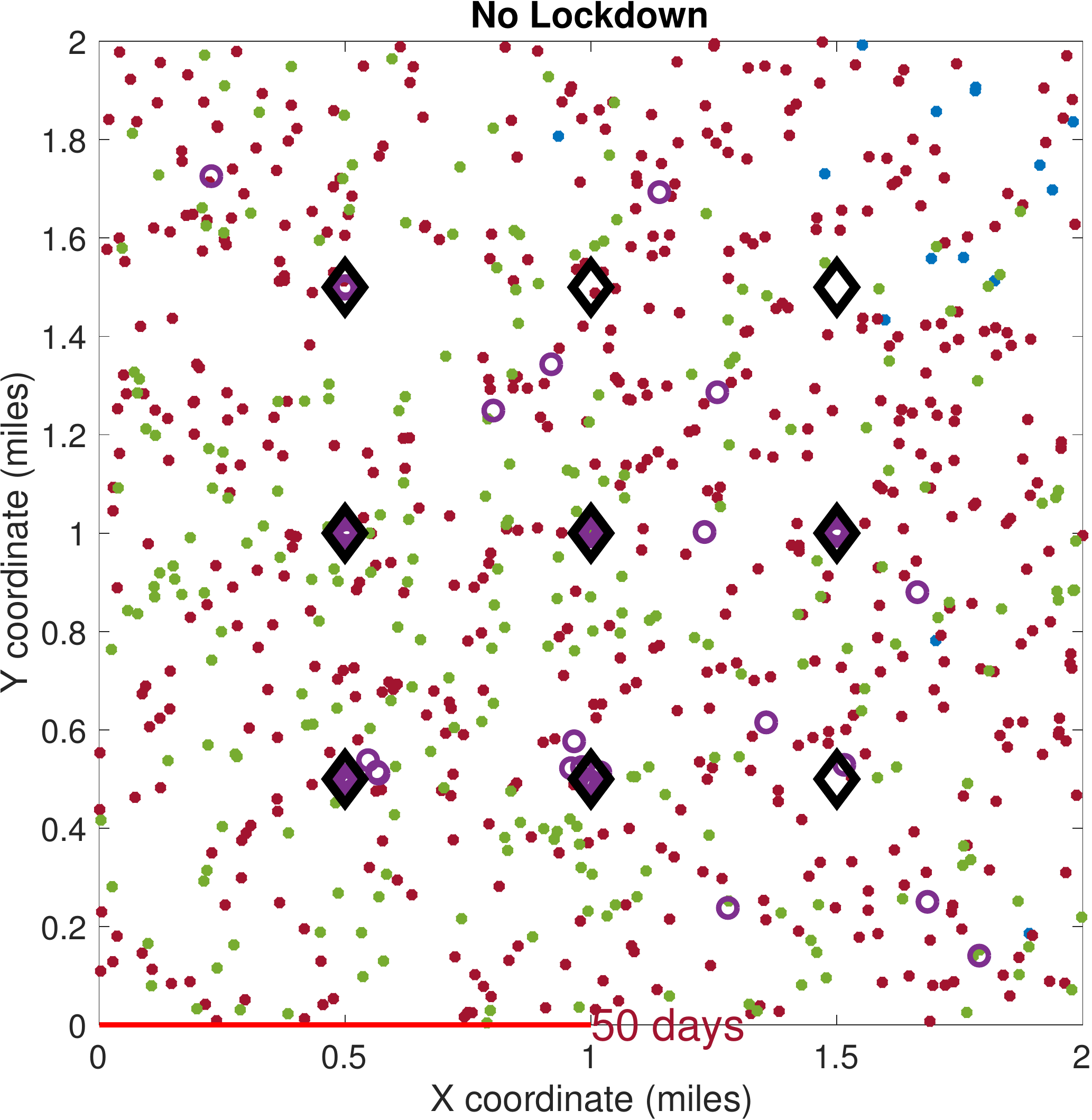} &
&\hspace*{-1.5in} \includegraphics[height=1.5in]{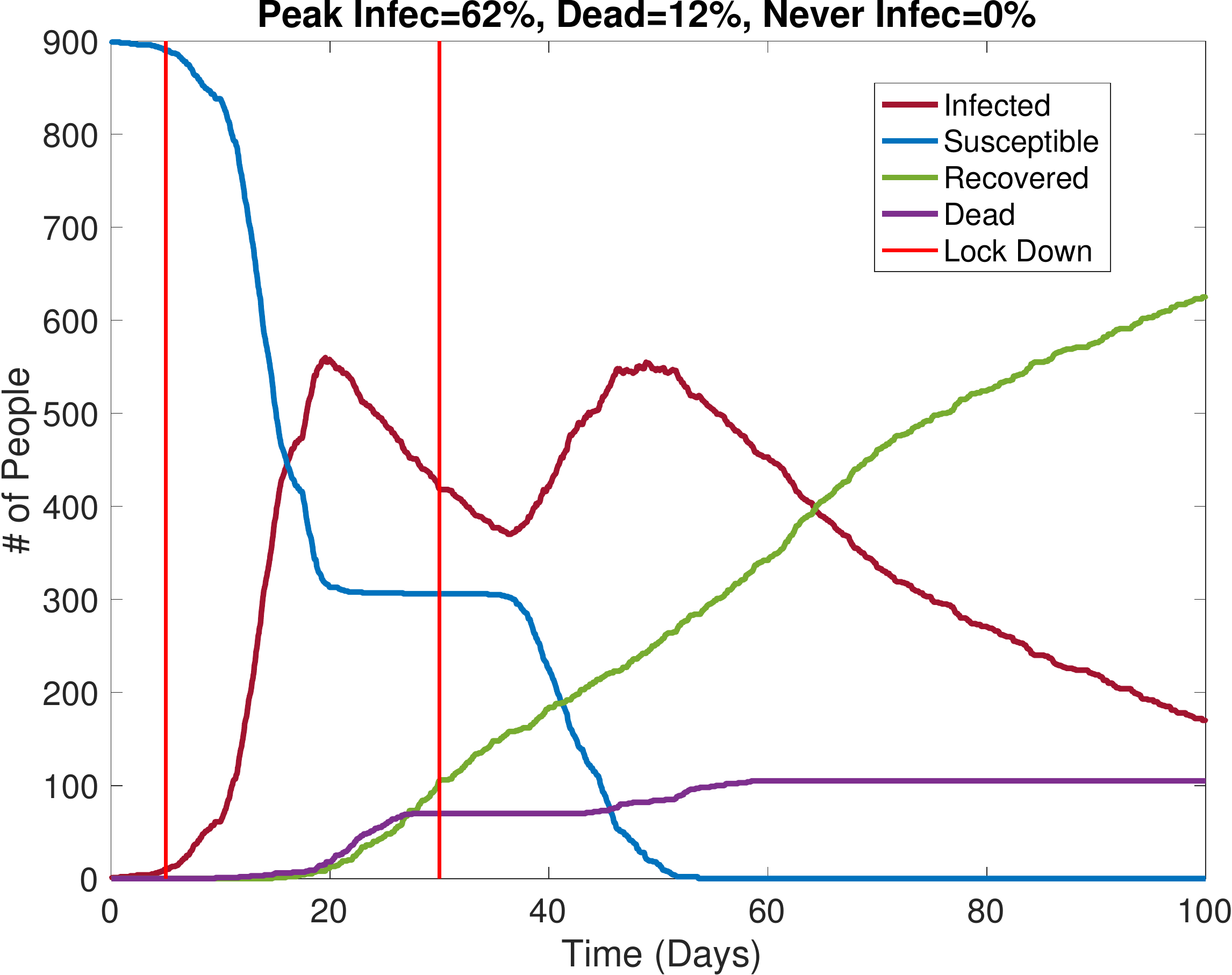}\\
\end{tabular}
\caption{Example 3: Model outputs when the restrictive measures are imposed on day 5 and removed 
on day 30.}
\label{examp3}
\end{center}
\end{figure}

\end{enumerate}

\subsection{Computational Complexity}

Since computational efficiency of the simulator is of vital 
important, we next 
study the computational cost of running ALPS for different variable sizes. In these experiments, we 
note the time taken by ALPS code on a McBook Pro 
laptop with Intel 2.8 GHz Core i7 processor and 16GB memory. 
The computational expense for a number of 
combination of $N$, $h$, and $T$ are given in the table below. 
Recall that in this implementation, the number of households units $h$ is a perfect square and the 
number of agents $N$ is a multiple (or close to it) of $h$. 
\begin{center}
\begin{tabular}{|c|c|c|c|}
\hline
$N$ (\# of Agents) & $h$ (\# of Households) & $T$ (Total days) & Time (sec) \\
\hline
90 & 9 & 100 & 1.526  \\
\hline
90 & 9 & 200 & 2.504  \\
\hline
180 & 9 & 100 & 2.173  \\
\hline
100 & 25 & 100 & 1.925  \\
\hline
486 & 81 & 100 & 6.680  \\
\hline
972 & 81 & 100 & 28.425  \\
\hline
1944 & 81 & 100 & 126.447  \\
\hline
968 & 121 & 100 & 29.916  \\
\hline
968 & 121 & 200 & 58.585  \\
\hline
\end{tabular}
\end{center}

From these results we are see that the computational cost is linear in $T$
with slope 1.
For a change in $N$ the number of agents, while keeping other variables fixed, 
the change in the computational cost is linear also. However, the rate of change is 
2 for smaller values but increases to 4 for the larger values. 
The changes in computational cost due to changes in the number of households, 
with other variables held fixed, are minimal. 

\section{Analyzing Lockdown Measures}

There are several ways to utilize this model for prediction, planning and 
decision making. We illustrate some of these 
ideas using examples. At first we show individual simulations under different scenarios and then
present results on average behavior obtained using hundreds of simulations. 
In these illustrations, we have used $N=972$ agents and $h = 81$ households.
\\

\begin{figure}[h]
\begin{center}
\begin{tabular}{cc}
\includegraphics[height=1.2in]{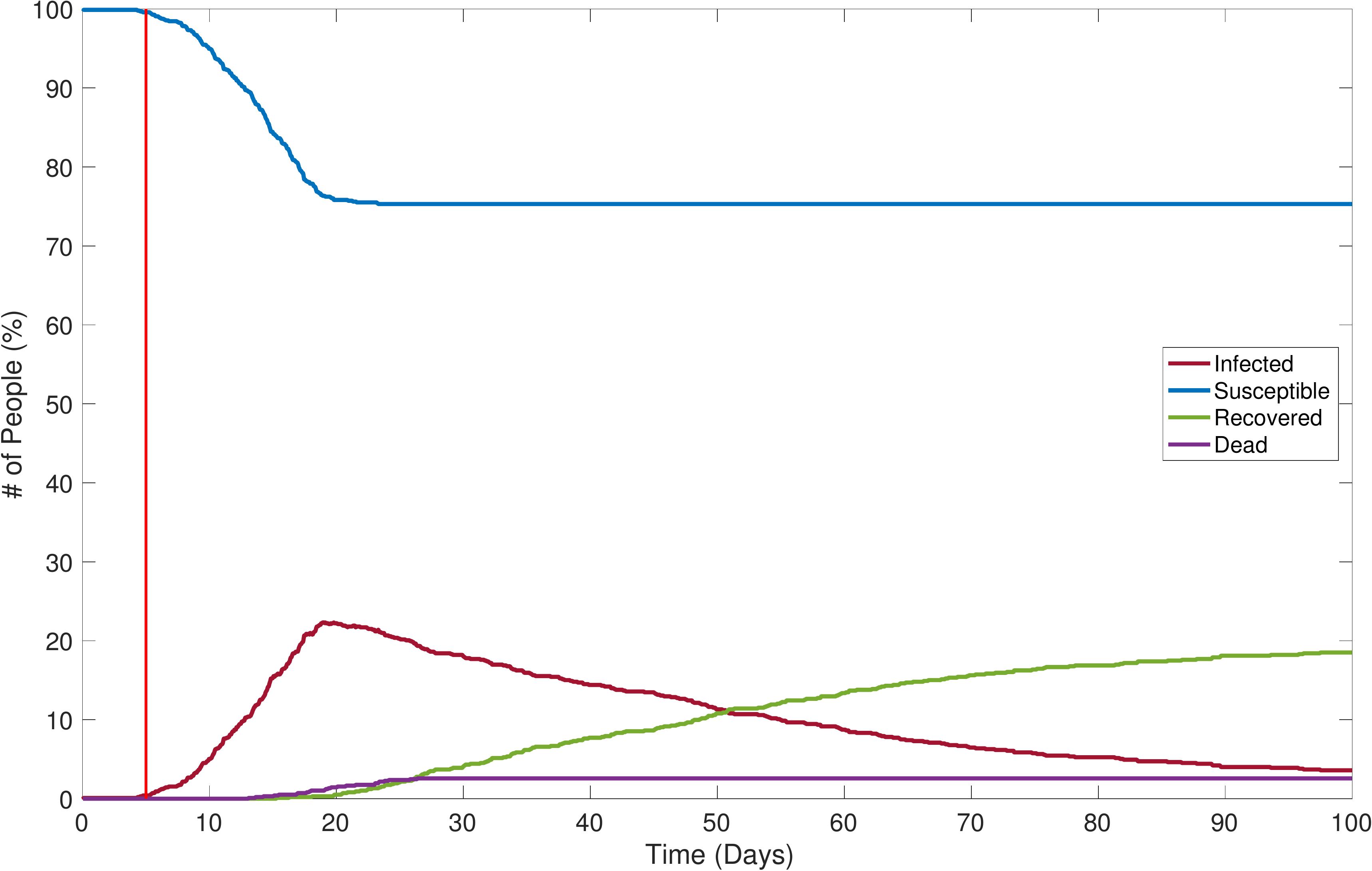}&
\includegraphics[height=1.2in]{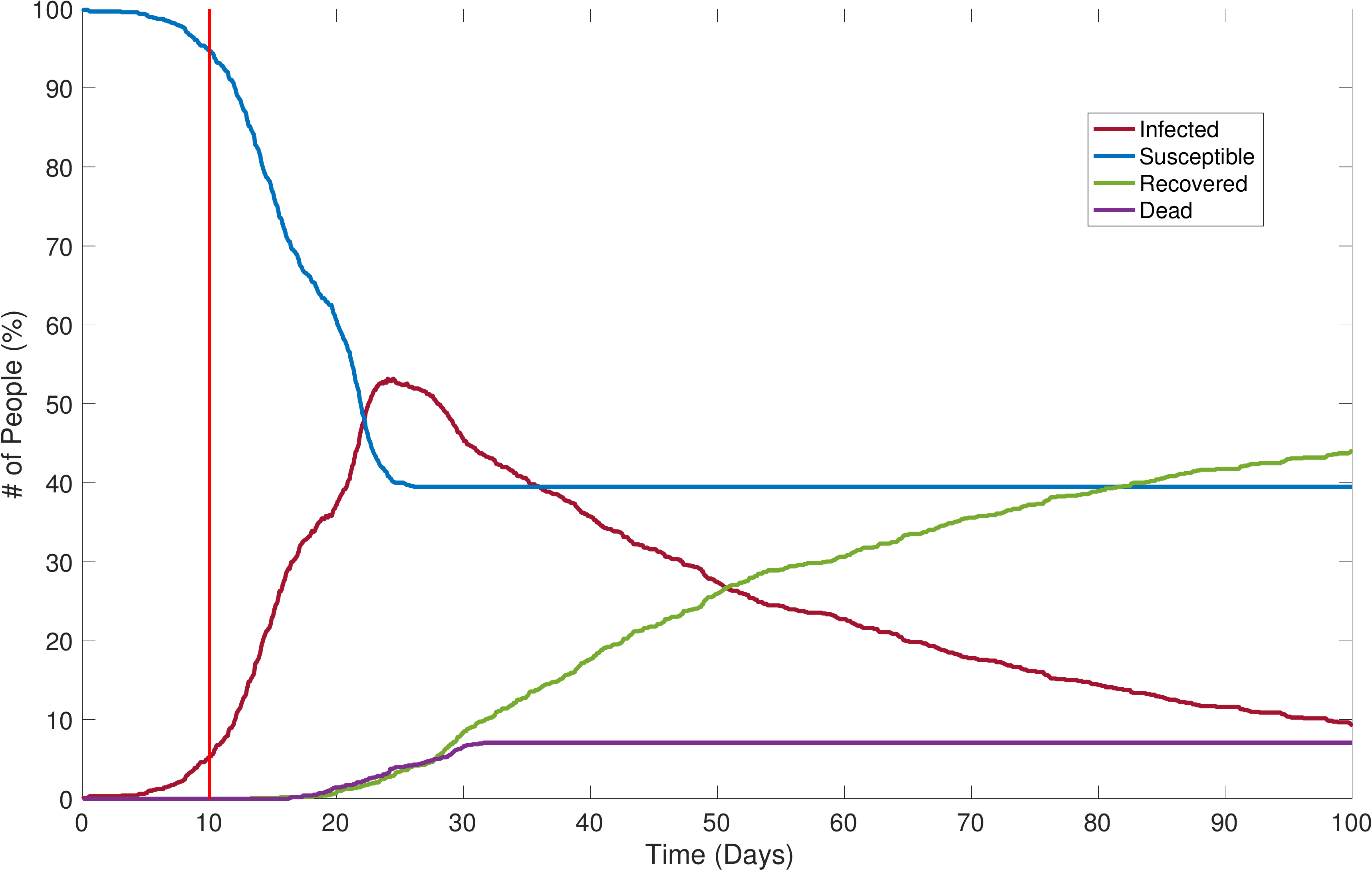}\\
\includegraphics[height=1.2in]{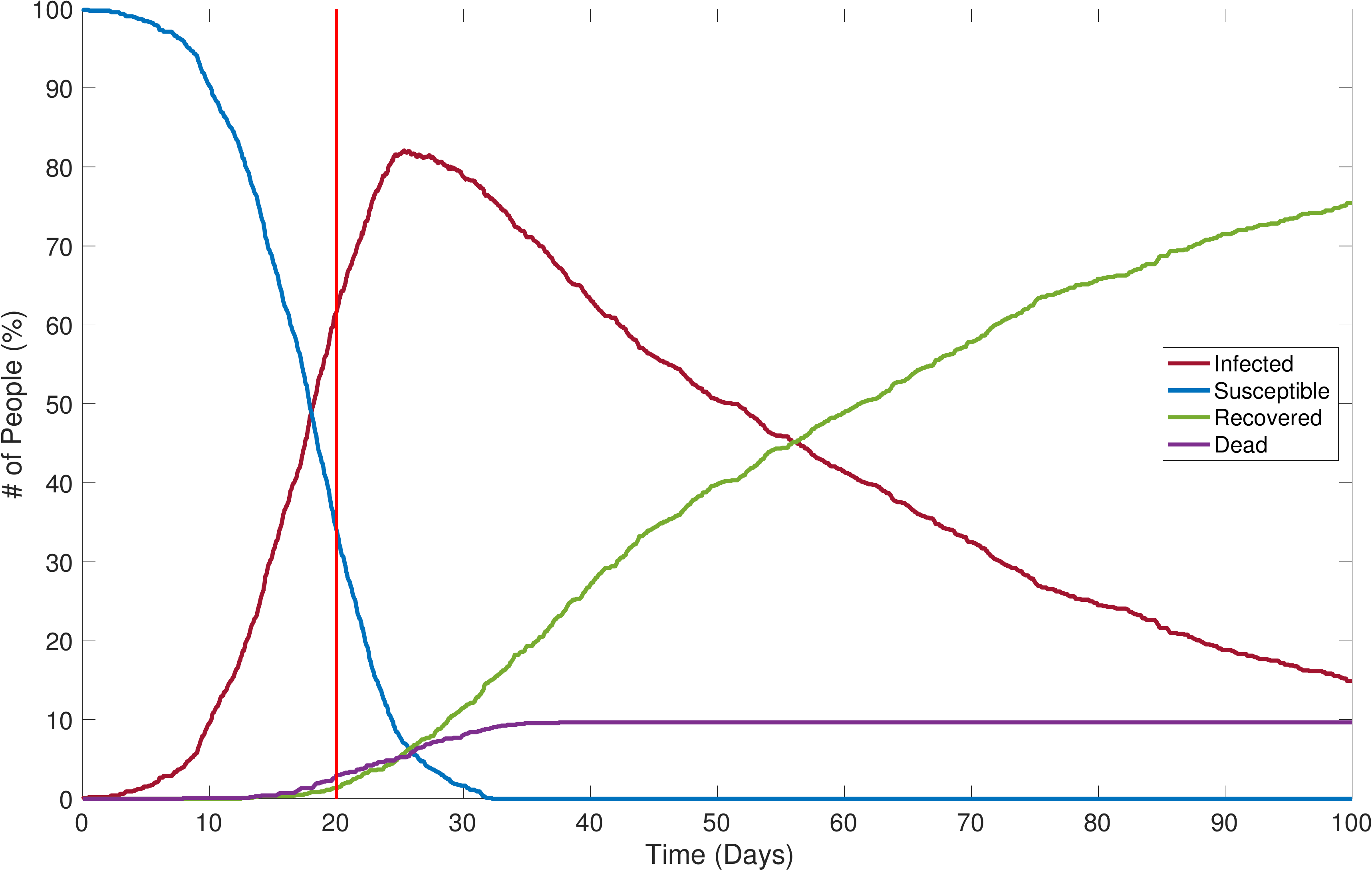} &
\includegraphics[height=1.2in]{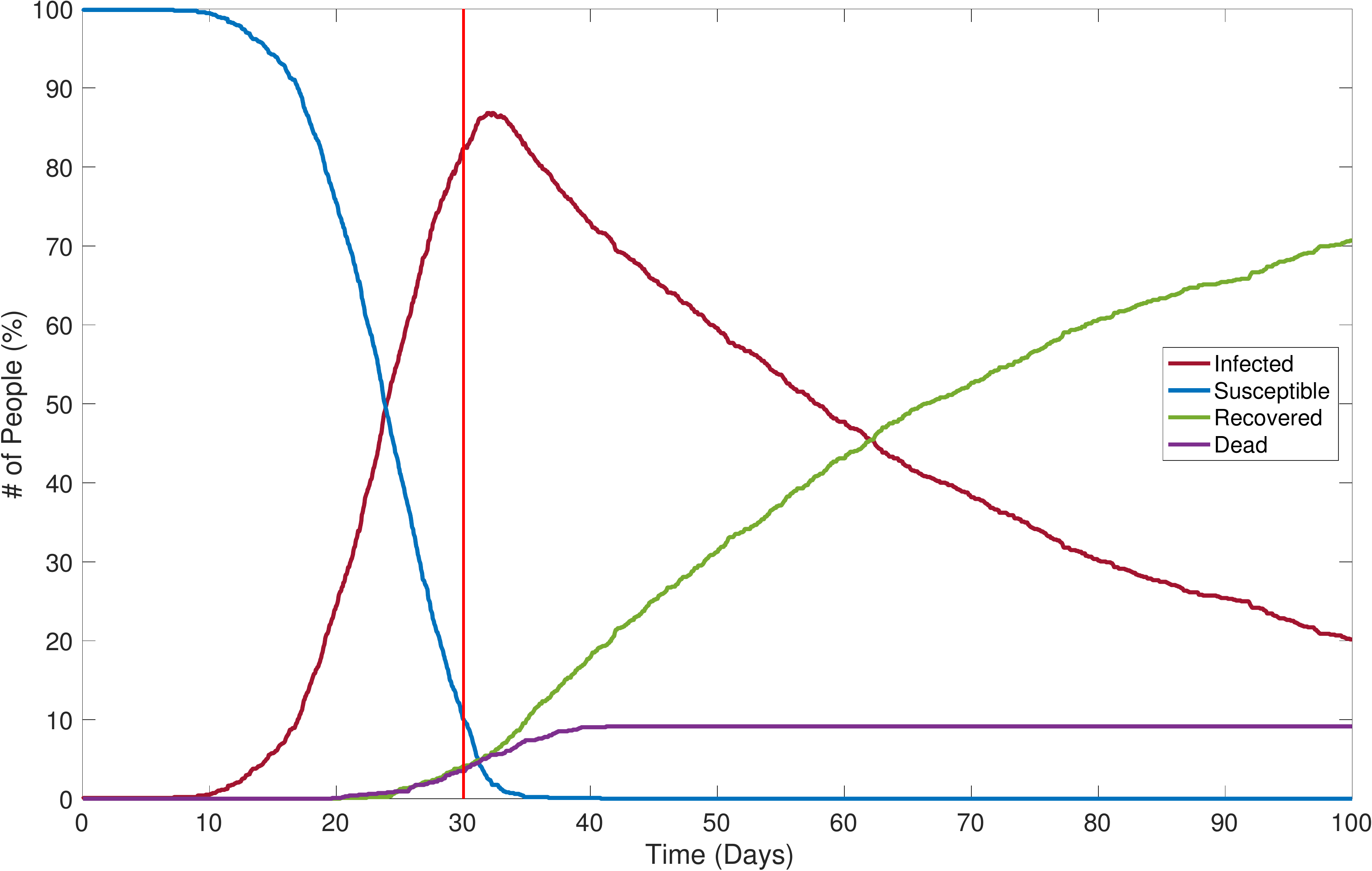}\\
\end{tabular}
\caption{Results from ALPS runs for different lockdown times. From top left to 
bottom right, the lockdowns are imposed on day 5, 10, 20, and 30.} \label{fig:study1}
\end{center}
\end{figure}

\noindent {\bf Effect of Timing of Imposition of Restrictions}: 
Fig.~\ref{fig:study1} shows some examples of ALPS outputs when a lockdown is imposed
on the community but at different times. From top-left to bottom-right, the plots show lockdowns starting on 
day 1, day 5, day 10, and day 20, respectively. Once the restrictions are imposed, they are not removed in these
examples. The best results are obtained for the earliest imposition of restrictions. In the top-left case, 
the peak infection rises to 22\% of the community, on day 18, and then comes down steadily. The fraction of 
fatalities is 2\% and the fraction of community never infected is 77\%. 
In case the restrictions are imposed on day 5, and all other parameters held same, there is a small 
change in the situation. The peak infection rises to 55\%, the fatalities increase to 7\% and 
the fraction of uninfected goes down to 40\%. We can see that an early imposition of lockdown measures
also helps reduce {\it peak infection} rates in the community. 

Sometimes we notice a saw-tooth shape in the curve for 
infected people. This implies that the even the small 
portion of mobile agents can break through and spread infections at other home 
units despite full restrictions being in place. This saw-tooth shape underscores the need for severely 
limiting mobility.
Even a small fraction of population being mobile can spread infections 
to the immobile agents and cause infections. 

The bottom two panels in Fig.~\ref{fig:study1} show results for a delayed lockdown, 
with the restrictions being imposed on day 20 and day 30. One can see that the peak infection
rate becomes quite high (82-84\%) and casualties mount to 10-11\%. The fraction of uninfected 
population falls to 
0\% in these cases. This shows that, under the chosen parameter settings,
day 20 is quite late in imposing lockdown conditions on the community, and the results are very 
similar to any later imposition. If the restrictions are imposed after 15-20 days, then there are no
uninfected people left in the community in a typical run of ALPS. 
\\ 

\begin{figure}
\begin{center}
\begin{tabular}{cc}
\includegraphics[height=1.2in]{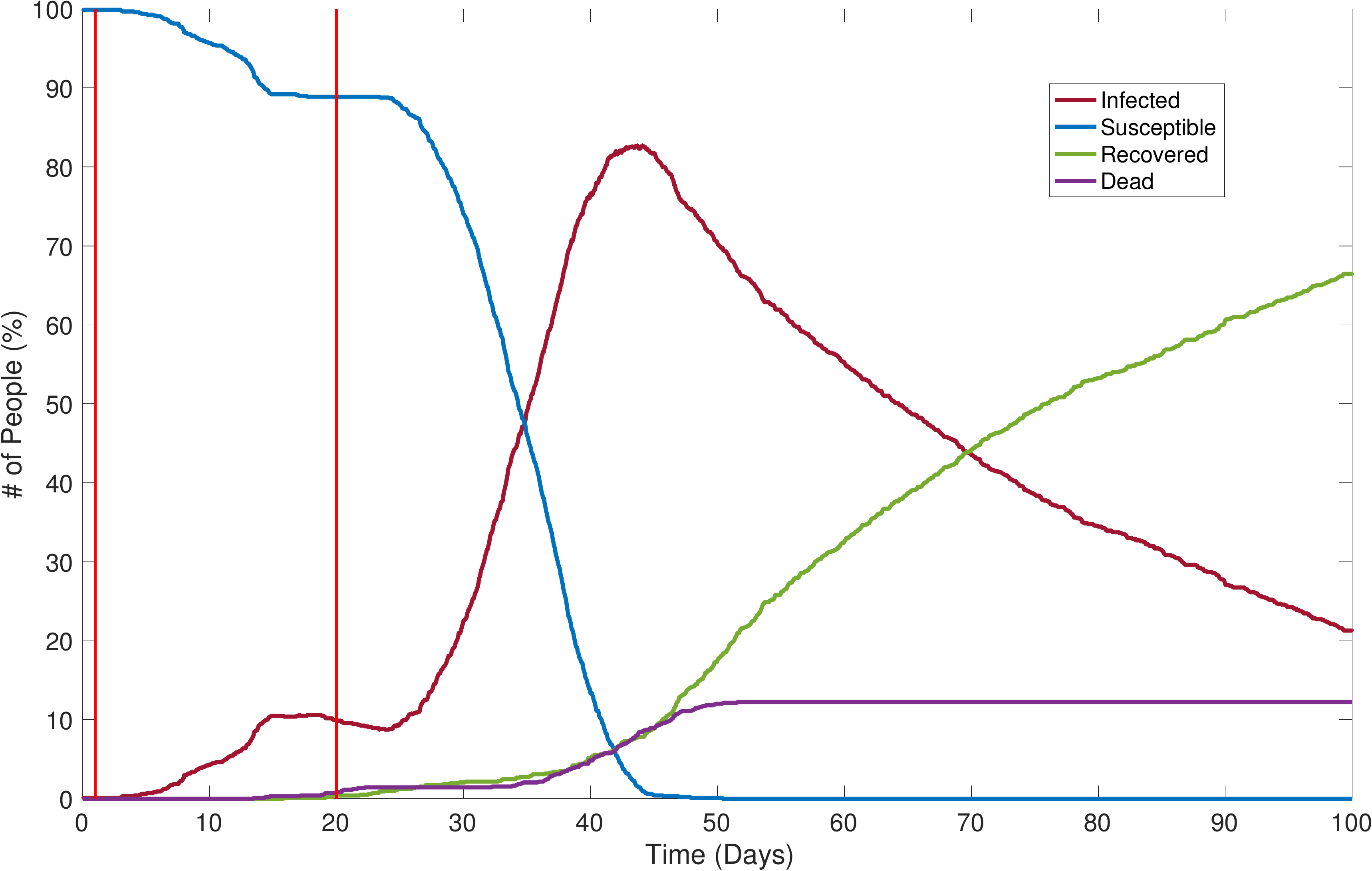} &
\includegraphics[height=1.2in]{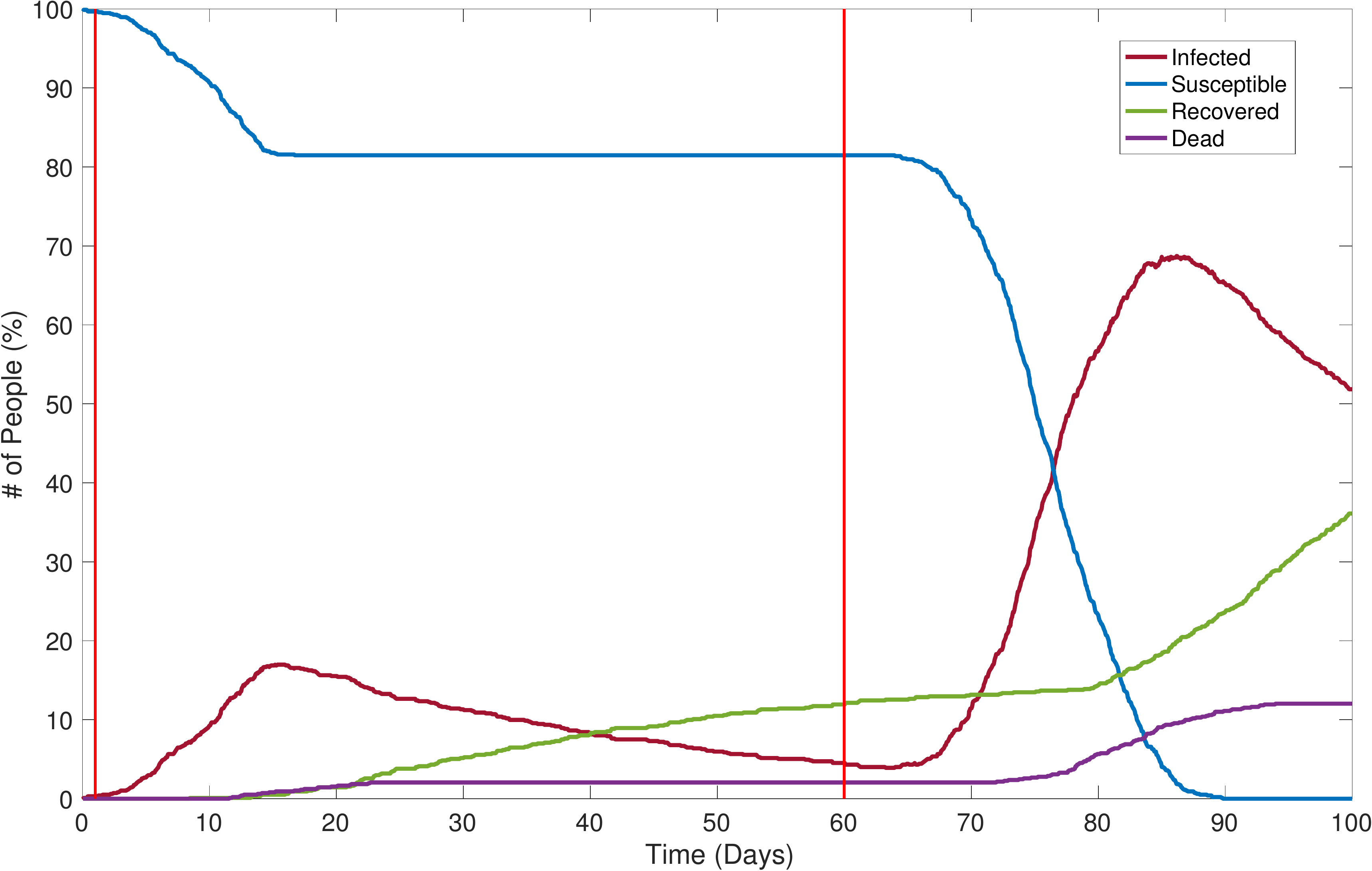} \\
$T_0 = 1$, $T_1 = 20$  & $T_0 = 1$, $T_1 = 60$ \\ 
\includegraphics[height=1.2in]{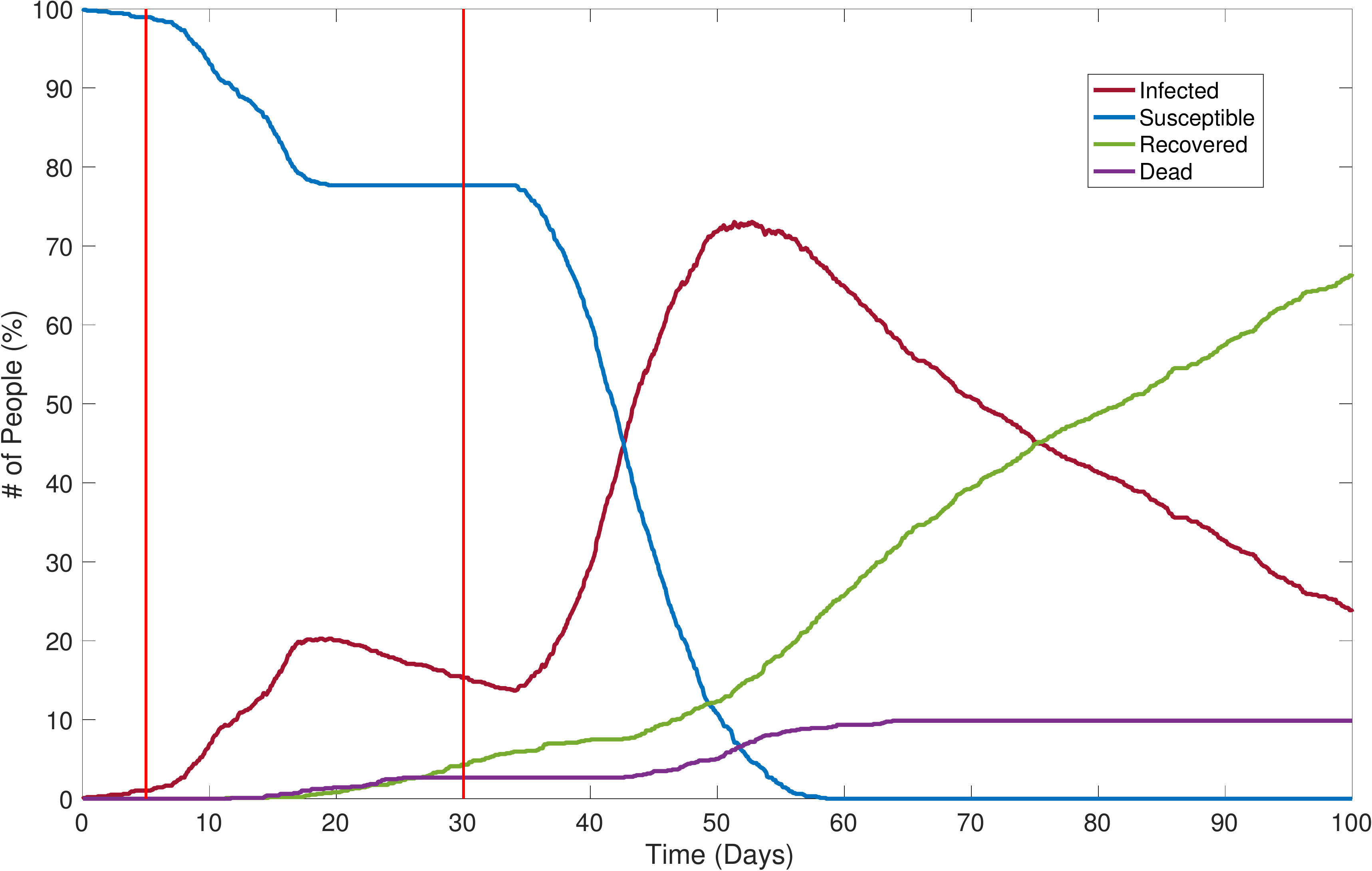} &
\includegraphics[height=1.2in]{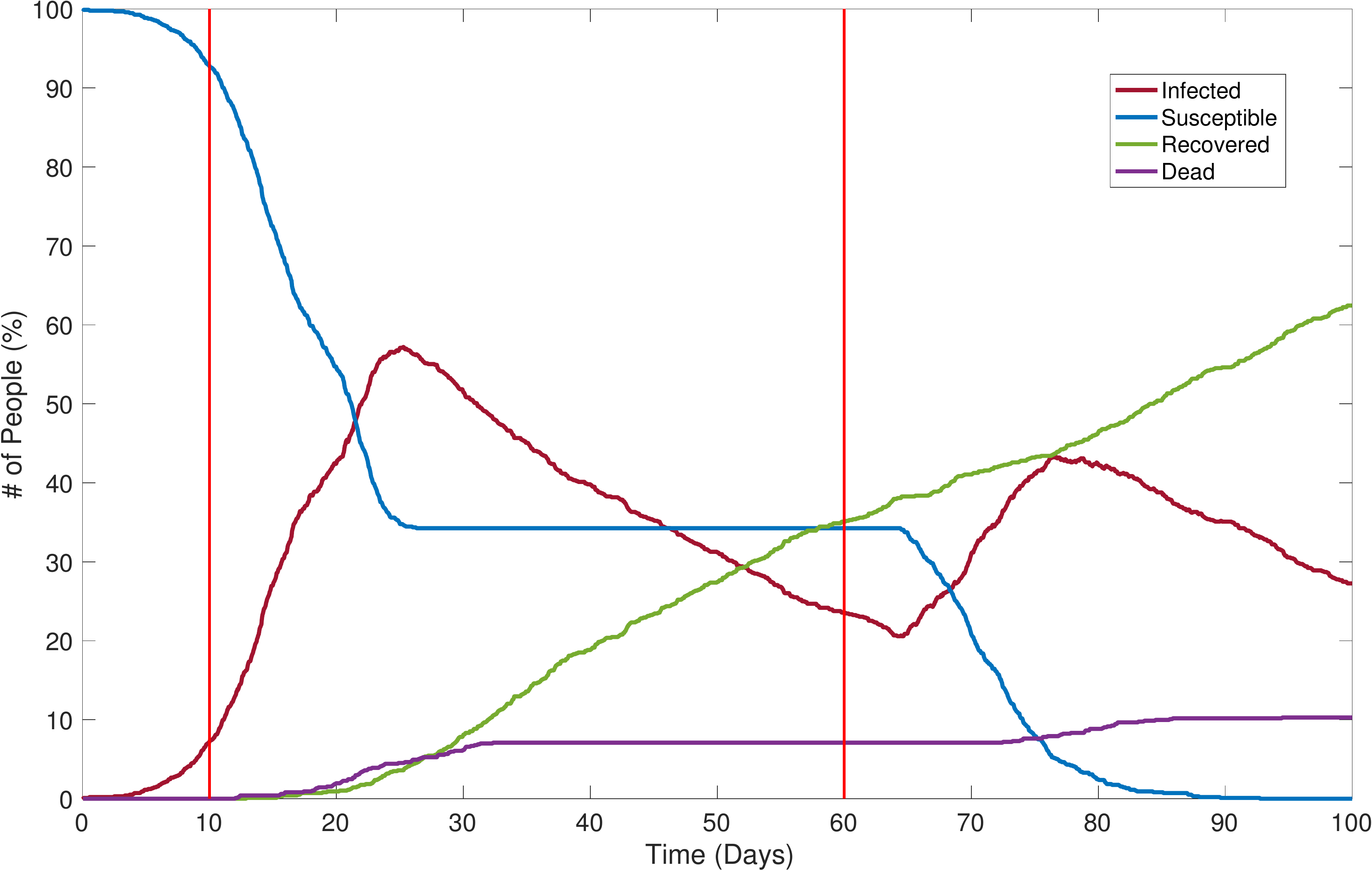} \\
$T_0 = 5$, $T_1 = 30$
& $T_0 = 10$, $T_1 = 60$
\end{tabular}
\caption{ALPS output for different combinations of impositions and lifting of 
restrictive conditions.} \label{fig:study2}
\end{center}
\end{figure}

\noindent {\bf Effect of Timing of Removal of Restrictions}: 
In the next set of simulations, we study the effects of lifting restrictions and thus
re-allowing full mobility in the community. Some sample results are shown in Fig.~\ref{fig:study2}. 
Each plot shows the evolution for a different starting time $T_0$ and the end time $T_1$. 
As these plots indicate, the gains made by early imposition of restrictions are nullified
when the restrictions are lifted. In all cases, after the lifting of restrictions, the full 
population gets infected eventually. 
Since we do no assume any change in the immunity levels of the agents over time, the 
results from lifting of restrictions are similar to not imposing any restrictions in the first place. The results 
appear to be same except for being shifted in time.  
\\

\begin{figure}
\begin{center}
\begin{tabular}{ccc}
\includegraphics[height=1.5in]{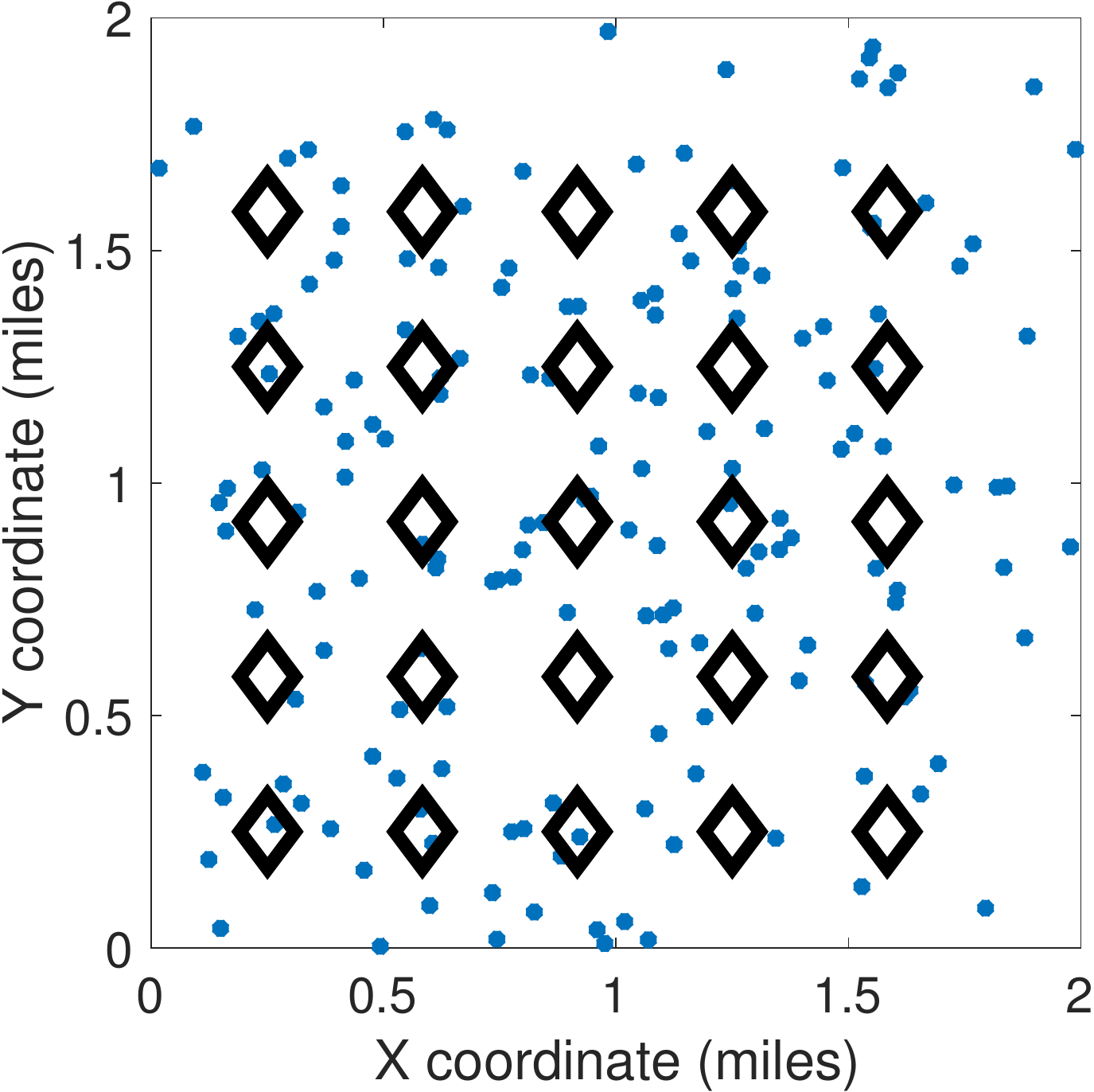}&
\includegraphics[height=1.5in]{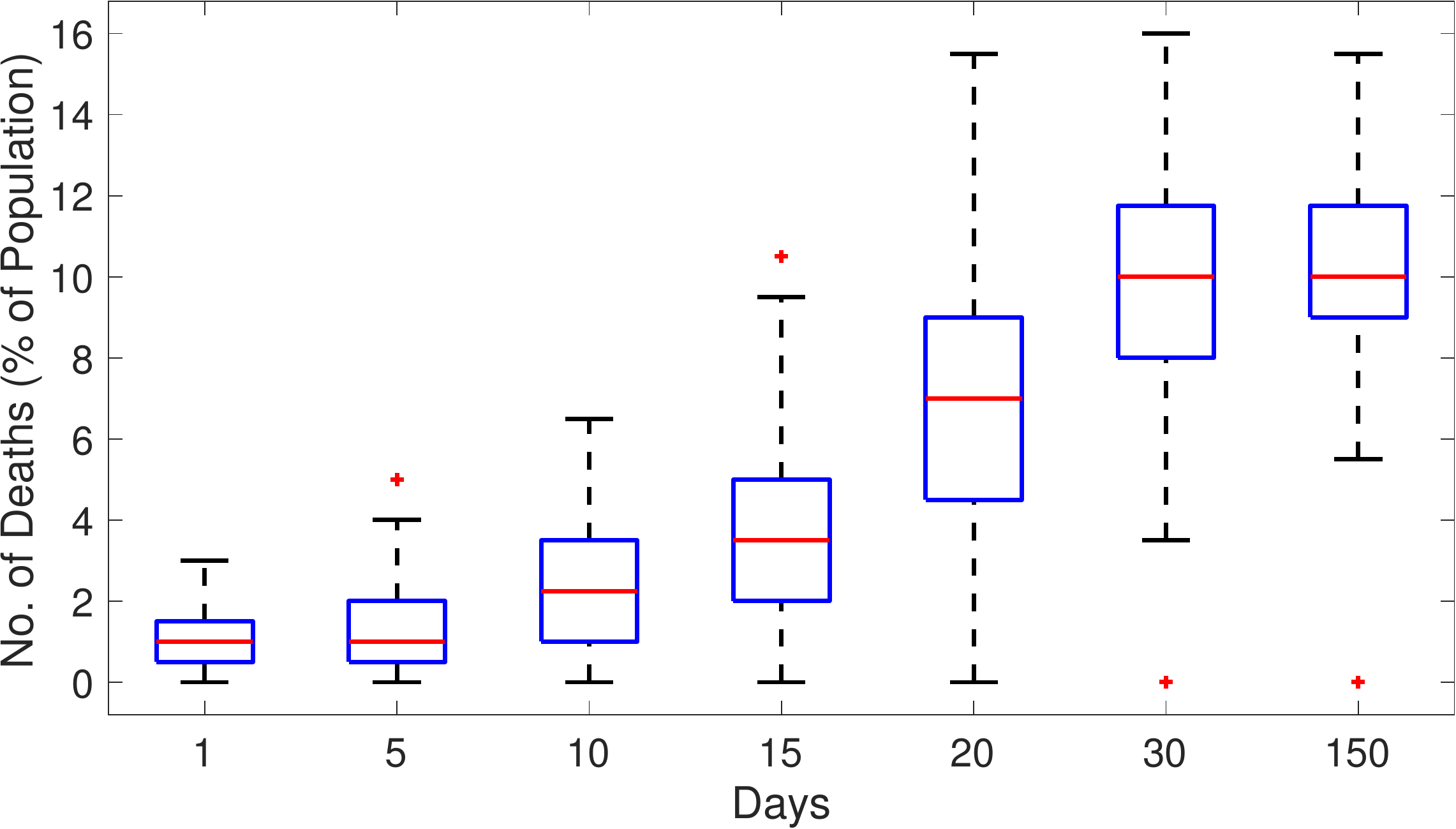}\\
\includegraphics[height=1.6in]{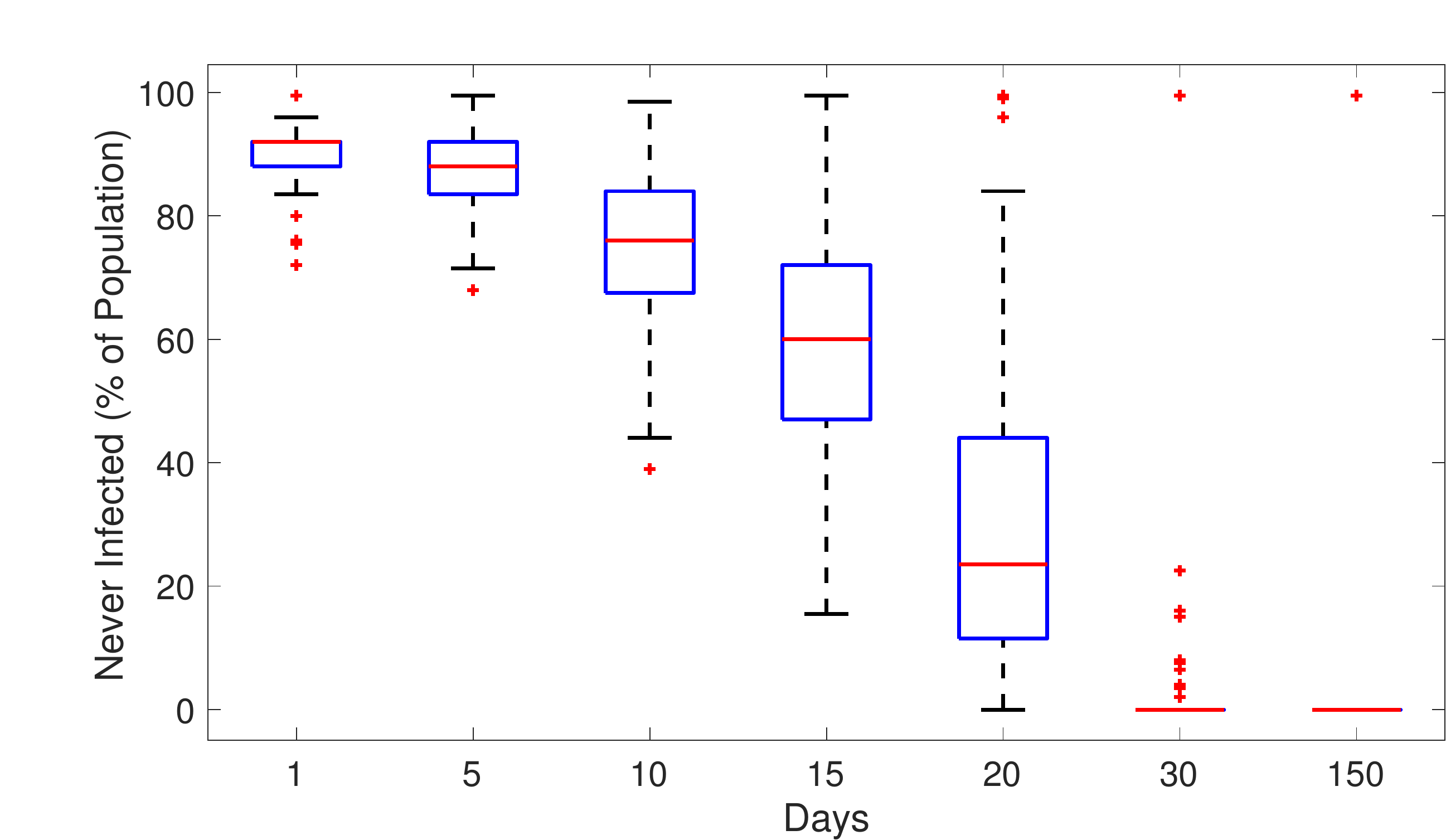} &
\includegraphics[height=1.5in]{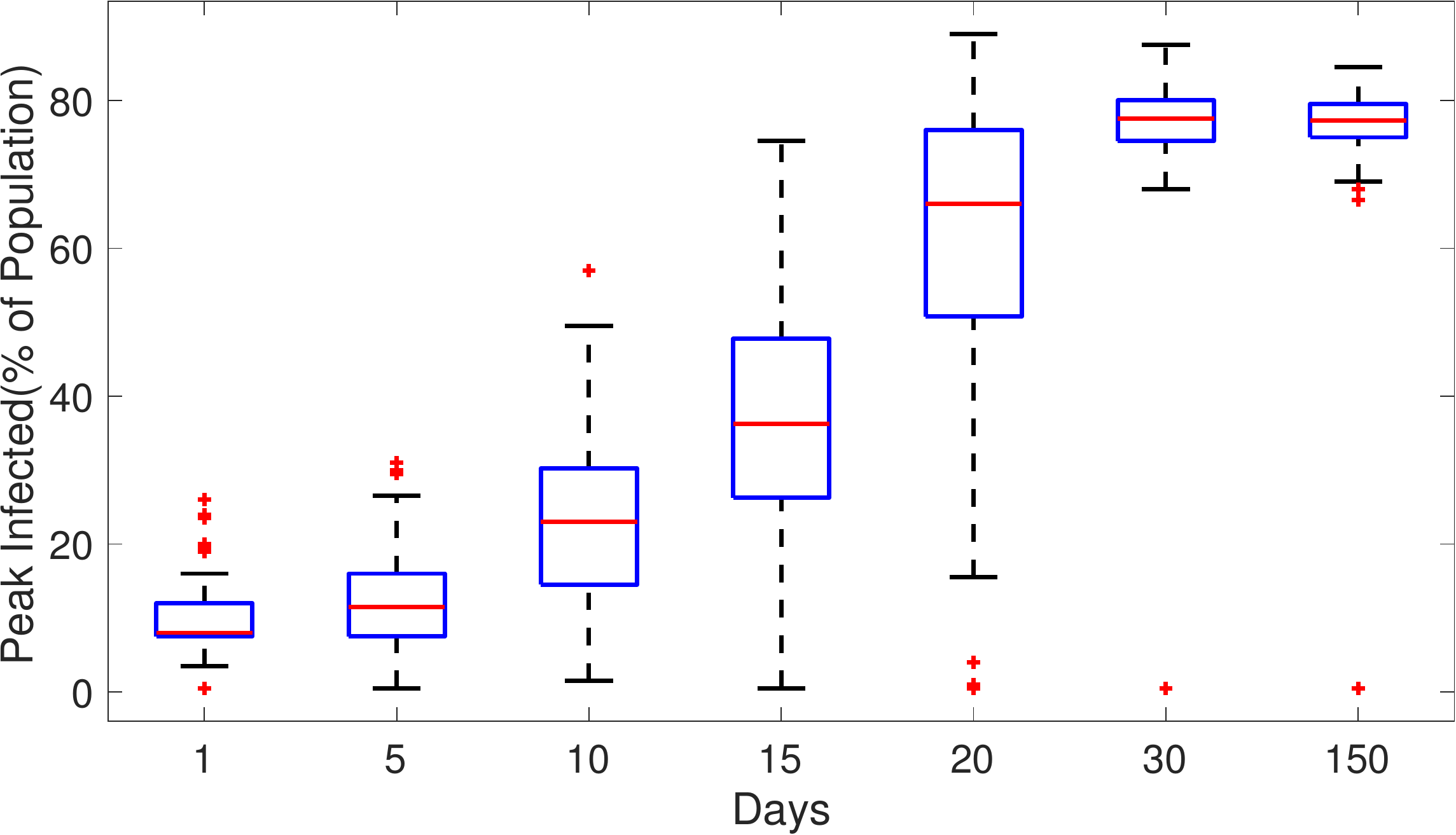}
\end{tabular}
\caption{Statistical summaries of infection variables obtained using 100 runs of ALPS, 
plotted against the starting day of the restrictions.} \label{fig:avg1}
\end{center}
\end{figure}

\noindent {\bf Statistical Summaries}: \\
In the next set of experiments, we compute average values of some variables of interest
using multiple ($=100$) runs of ALPS. In the first result, we study three variables -- 
number of deaths, number of people remaining uninfected, and the peak infection 
rate -- using $N=200$ agents living in a community of $h=25$ households, observed over 
$[0,150]$ days. The top left panel of Fig.~\ref{fig:avg1}  shows an example of the configuration with 
25 households and 200 agents in the community. 
We vary the start time $T_0$ (start day of restrictions) from $1$
to $30$ and then to $150$ and study the resulting 
outcomes using 100 runes of ALPS. (The value
of $T_0 = 150$  implies that the restrictions are never
imposed in that setting.) The remaining panels in Fig.~\ref{fig:avg1} show box plots of the
three variables changing with $T_0$. 
\begin{itemize}
\item {\bf Death Rate}: Top right shows the percentage fatalities 
increasing from around 1\% to almost 11\% as $T_0$ changes from $1$ to $30$.  
the largest rate of increase is observed when $T_0$ is between $10$ to $30$ days. 

\item {\bf Number of Uninfected}: In the bottom left panel we see a decrease in the number of 
uninfected population from around 90\% to 0\% as $T_0$ increases. In case the restrictions are
not imposed in first 30 days after the first infection, there is no agent left uninfected in the 
community. 
 
\item {\bf Peak Infection Rate}: In case the restrictions are imposed on the first day after the 
infection, the peak infection rate is contained to be $10\%$. As $T_0$ is increased and the 
restrictions are delayed the peak infection rate rises to almost $80\%$ of the community. 

\end{itemize}

\begin{figure}
\begin{center}
\begin{tabular}{cc}
\includegraphics[height=1.5in]{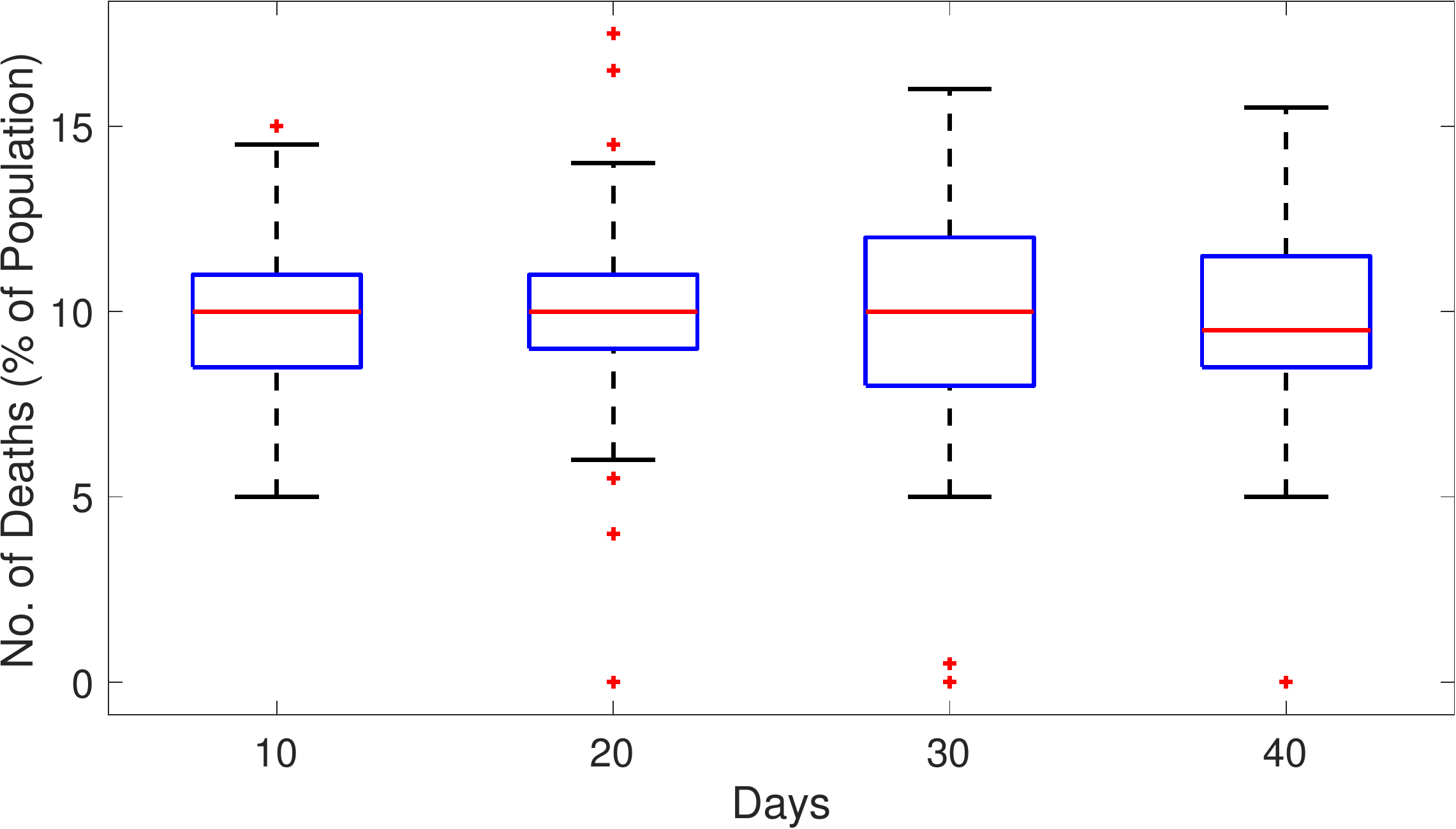}&
\includegraphics[height=1.5in]{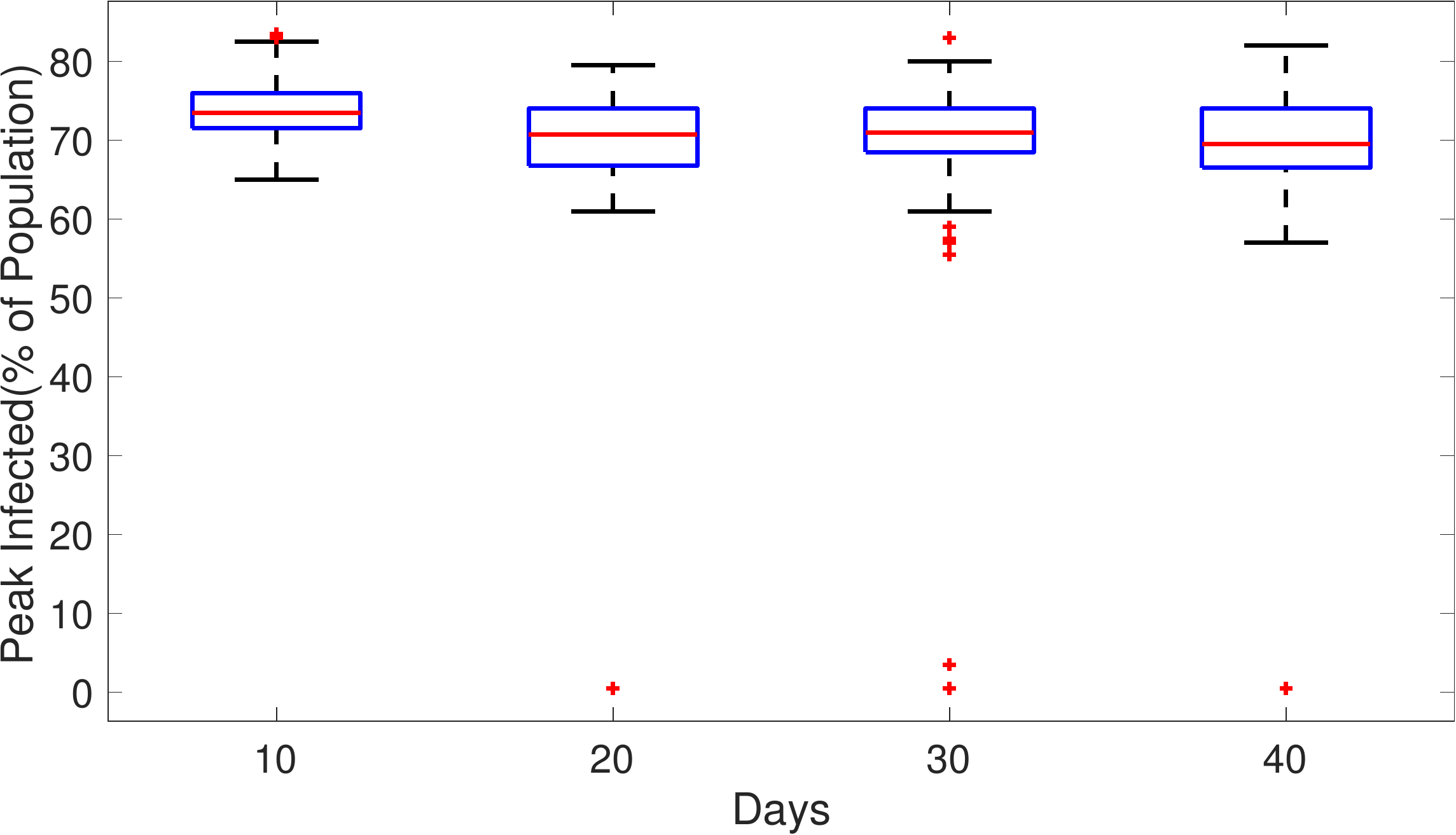}\\
\end{tabular}
\caption{Statistical summaries of infection variables obtained using 100 runs of ALPS, 
plotted against the reopening day.} \label{fig:avg2}
\end{center}
\end{figure}

In Fig.~\ref{fig:avg2} we study the effect of changing $T_1$ while $T_0 = 1$ is kept 
fixed (and other experimental conditions being same as in the last experiment). The 
results show that there is no difference in the eventual number of deaths and the peak infection 
rates when $T_1$ is changed from $10$ to $40$. This is because agent immunity 
and other infection factors are kept constant over time, and lifting of restrictions gives
results akin to results from unrestricted conditions, irrespective of $T_1$. 

\section*{Discussion}

The strengths and limitations of ALPS model  are the following. 
ALPS provides an efficient yet comprehensive modeling of the spread of infections in 
a self-contained community, using simple model assumptions. The model can prove
very useful in evaluating costs and effects of imposing social lockdown measures
in a society. 

In the current version, the initial placement of agents is set to be normally 
distributed with means given by their home units
and fixed variance. This variance is kept large to allow for near arbitrary placements of 
agents in the community. In practice,  however,  agents typically follow semi-rigid daily schedules of being 
at work, performing chores, or being at home. Thus, at the time of imposition  
of a lockdown, the agents can be better placed in the scenes according to their 
regular schedules rather than being placed arbitrarily. 

In terms of future directions, 
there are many ways to develop this simulation model to capture more realistic 
scenarios: (1) It is possible to model multiple, interactive communities instead of a single 
isolated community. (2) One can 
include typical daily schedules for agents in the simulations. A typical agent 
may leave home in the morning, spent time in the office during the day, and 
return to home in the evening. (3) It is possible to provide an age demographics to the 
community and assign immunity to agents according to their demographic labels~\cite{chang-etal:2020}. 
(4) As more data becomes available in the future, one can change immunity levels of 
agents over time according to the spread and seasons. (5) In practice, when an agent is infected, 
he/she goes  through different stages of the disease, associated 
with varying degrees of mobility
\cite{perez-etal-2009}. 
One can introduce an additional variable to track these stages of infections in the model and change agent 
mobility accordingly.

\section*{Conclusion}
This paper develops an agent-based simulation model, called ALPS,  for modeling spread of 
an infectious disease in a closed community. A number of simplifying and 
reasonable assumptions 
makes the ALPS efficient and effective for statistical analysis. The model is validated
at a population level by comparing with the popular SIR model in epidemiology.
These results indicate that: (1) Early imposition of lockdown measures (right after the first infection)
significantly reduce infection rates and fatalities; (2) Lifting of lockdown measures recommences the  
spread of the disease and the infections eventually reach the same level as the 
unrestricted community; (3) In absence of any extraneous solutions (a medical treatment/cure, a weakening 
mutation of the virus, or a natural development of agent immunity), the only viable option for preventing
large infections is the judicious use of lockdown measures. 

\section*{Acknowledgements}
This research was supported in part by the grants NSF DMS-1621787 and NSF DMS-1953087.

\appendix

\section{Listing of ALPS parameters}
Table 1 provides a listing of all the parameters one can adjust in 
ALPS to achieve different scenarios. It also provides some typical values used in the 
experiments presented in the paper. 

\begin{table}[h]
\begin{center} \label{tab:ALPS-params}
\begin{tabular}{|c|c|c|c|c|}
\hline
No. & Sym & Explanation & Range &  Typical Values\\
   \hline
\multicolumn{5}{|c|}{\bf Community Parameters} \\
\hline
 1  & $t$ & Current time instance & $0 \leq t \leq T$ & $T = 4800$ hours \\
   \hline
2     & $N$ & Total number of agents &  $N > 1$&   $N \approx1000$ \\
   \hline
 3    & $h$ & Total number of housing units & $h > 0$ &   $h = 9, 49, 81$ \\
   \hline \hline
\multicolumn{5}{|c|}{\bf Motion Related Parameters} \\
\hline
1 & $\alpha$ & Rate at which a person heads home & $\alpha \in \real_+$ & $\alpha = 0.2$ \\
\hline
2 & $\mu$ & Relative proportion of homeward motion  & $\mu \in [0,1]$ & $\mu = 0$ -- lockdown  \\
& & and random walk &  & $\mu = 1$ -- free  \\
\hline
3 & $\sigma$ & Variance in acceleration noise & $\sigma \in \real_+$ & $\sigma = 0.0001$ mph \\
\hline
\hline
\multicolumn{5}{|c|}{\bf Social Distancing Related Parameters} \\
\hline
1 & $\rho_0$ & Fraction of people following restrictions & $\rho_0 \in [0,1]$ &  $\rho = 0.98$ \\
\hline
2 & $T_0$ & when lockdown starts & $T_0 \in \{1,2,\dots,\}$ & $T_0 = 5$ days \\
\hline
3& $T_1$ & when lockdown ends & $T_1 \in \{1,2,\dots,\}$ & $T_1 = 30$ days \\
\hline \hline
\multicolumn{5}{|c|}{\bf Infection Related Parameters} \\
\hline
1 & $r_0$ & Max distance to catch infection & $r_0 \in \real_+$ & $r_0 = 6$ feet\\
\hline
2 & $\tau_0 $ & Min. exposure time to catch infection & $\tau_0 \in \real_+$ & $\tau_0 = 5$ hours \\
\hline
3 & $p_I$ & Prob. of infection at each time & $p_I \in [0,1]$ & $p_I = 0.01$ \\
\hline \hline
\multicolumn{5}{|c|}{\bf Recovery/Death Related Parameters} \\
\hline
1 & D & Disease type -- fatal or non-fatal & D = FT, NFT  & \\
\hline
2 & $p_F$ & Prob. of FT once infected & $p_F \in [0,1]$  & $p_F = 0.1$ \\
\hline
3 & $T_R$ & Period before recovery starts for NFT & $T_R \geq 0$  & $T_R = 7$  days\\
\hline
4 & $T_D$ & Period before death can occur for FT & $T_D \geq 0$  & $T_D = 7$  days\\
\hline
5 & $p_D$ & Prob. of death after $T_D$ at each $t$ & $p_D \in [0,1]$  & $p_D = 0.1$ \\
\hline
6 & $p_R$ & Prob. of recovery after $T_R$ at each $t$ & $p_R \in [0,1]$  & $p_R = 0.001$ \\
\hline
\end{tabular}
\caption{Listing of parameters associated with different model components of ALPS.}
\end{center}
\end{table}


\bibliography{Biblio}

\end{document}